%% file: programmable.tex
\documentclass[aps,pra,twocolumn,superscriptaddress]{revtex4-1} 

\input{AuxTexFiles/preamble}

\begin{document}
\begin{bibunit}
\renewcommand{\figurename}{\textbf{Fig.}}
\title{Programmable Interactions and Emergent Geometry in an Atomic Array}

\author{Avikar Periwal}
\author{Eric S. Cooper}
\author{Philipp Kunkel}
\author{Julian F. Wienand}
\author{Emily J. Davis}
\affiliation{Department of Physics, Stanford University, Stanford, California 94305, USA}
\author{Monika Schleier-Smith}
\affiliation{Department of Physics, Stanford University, Stanford, California 94305, USA}
\affiliation{SLAC National Accelerator Laboratory, Menlo Park, CA 94025}

\begin{abstract}
Interactions govern the flow of information and the formation of correlations in quantum systems, dictating the phases of matter found in nature and the forms of entanglement generated in the laboratory. Typical interactions decay with distance and thus produce a network of connectivity governed by geometry, e.g., by the crystalline structure of a material or the trapping sites of atoms in a quantum simulator~\cite{bloch2012quantum,browaeys2020many}.  However, many envisioned applications in quantum simulation and computation require richer coupling graphs including nonlocal interactions, which notably feature in mappings of hard optimization problems onto frustrated spin systems~\cite{Das2008,gopalakrishnan2011frustration,strack2011dicke,mcmahon2016fully,berloff2017realizing} and in models of information scrambling in black holes~\cite{hayden2007black,maldacena2016remarks,bentsen2019treelike,belyansky2020minimal}. Here, we report on the realization of programmable nonlocal interactions in an array of atomic ensembles within an optical cavity, where photons carry information between distant atomic spins~\cite{leroux2010implementation,barontini2015deterministic,hosten2016measurement,welte2018photon,pedrozo2020entanglement,davis2019photon,davis2020protecting,muniz2020exploring}.  By programming the distance-dependence of interactions, we access effective geometries where the dimensionality, topology, and metric are entirely distinct from the physical arrangement of atoms.  As examples, we engineer an antiferromagnetic triangular ladder, a M\"{o}bius strip with sign-changing interactions, and a treelike geometry inspired by concepts of quantum gravity~\cite{barbon2013fast,gubser2017p,heydeman2016tensor,bentsen2019treelike}. The tree graph constitutes a toy model of holographic duality~\cite{gubser2017p,heydeman2016tensor}, where the quantum system may be viewed as lying on the boundary of a higher-dimensional geometry that emerges from measured spin correlations \cite{qi2018does}.  Our work opens broader prospects for simulating frustrated magnets and topological phases, investigating quantum optimization algorithms, and engineering new entangled resource states for sensing and computation.

\end{abstract}
\date{\today}

\maketitle

Driving matter with light offers a powerful approach to engineering quantum mechanical systems \cite{rudner2020band,kennedy2015observation,aidelsburger2015measuring,struck2013engineering,vaidya2018tunable,islam2013emergence,jurcevic2014quasiparticle,fausti2011light,wang2013observation,leroux2010implementation,barontini2015deterministic,bohnet2016quantum,hosten2016measurement,welte2018photon,pedrozo2020entanglement,davis2019photon,davis2020protecting,muniz2020exploring,leonard2017supersolid}.  In electronic materials \cite{fausti2011light,wang2013observation} and in artificial materials composed of trapped atoms in quantum simulators \cite{rudner2020band,kennedy2015observation,aidelsburger2015measuring,struck2013engineering,vaidya2018tunable}, optical driving allows for controlling transport properties, interactions, and correlations.  For atoms in optical cavities \cite{vaidya2018tunable} or trapped-ion qubits \cite{islam2013emergence,jurcevic2014quasiparticle}, photons or phonons can mediate interactions of tunable range.  Experiments have investigated the influence of the range of interactions on the growth of quantum correlations \cite{islam2013emergence,jurcevic2014quasiparticle} and harnessed infinite-range interactions to generate collective entangled states \cite{leroux2010implementation,barontini2015deterministic,bohnet2016quantum,hosten2016measurement,welte2018photon,pedrozo2020entanglement}.  All-to-all interactions mediated by light have furthermore enabled quantum simulations of phenomena ranging from supersolidity \cite{leonard2017supersolid} to dynamical phase transitions \cite{muniz2020exploring}.

Yet many objectives in quantum simulation and computation demand more versatile control of the graph of interactions \cite{gopalakrishnan2011frustration,strack2011dicke,hung2016quantum,bentsen2019treelike,belyansky2020minimal,ozeri2020quantum}.  Engineering a wider range of nonlocal coupling graphs opens prospects for simulating exotic frustrated magnets supporting spin-glass phases \cite{strack2011dicke,gopalakrishnan2011frustration} and topologically ordered states \cite{hung2016quantum}, implementing combinatorial optimization algorithms \cite{mcmahon2016fully,berloff2017realizing,marsh2021enhancing,anikeeva2020number}, and probing toy models of quantum gravity \cite{bentsen2019treelike,belyansky2020minimal,kollar2019hyperbolic}.  These goals have motivated proposals for programming the distance-dependence of spin-exchange interactions in arrays of atoms or ions by tailoring the frequency spectrum of a drive field \cite{hung2016quantum,bentsen2019treelike,ozeri2020quantum}, which couples the spins to a single mode of light or motion.  

\begin{figure*}
\includegraphics[width=\textwidth]{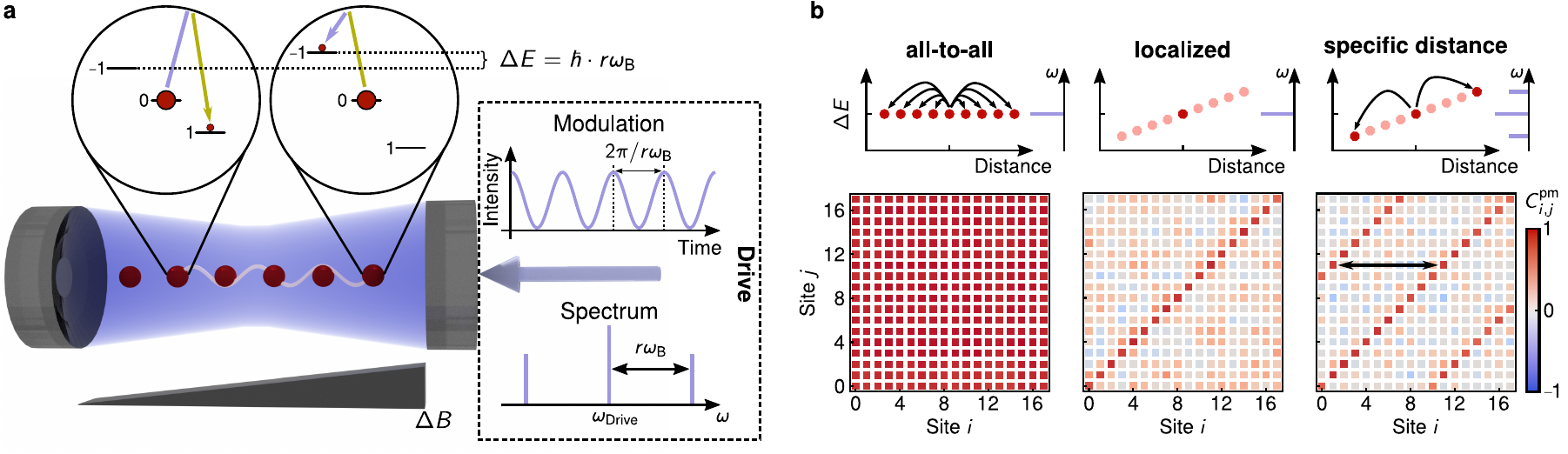}
\caption{\textbf{Engineering distance-dependent interactions.} 
\textbf{a} An array of atomic ensembles is trapped inside an optical cavity. We apply a magnetic field gradient $\Delta B$ in the longitudinal direction, which leads to a difference $\Delta E = \hbar r\omega_B$ in the Zeeman splittings of atoms separated by a distance $r$. By modulating the intensity of the drive field at a frequency $r\omega_{B}$ we generate correlated atom pairs in states $m = \pm1$ at distance $r$.
\textbf{b} Measured correlations $C^\text{pm}$ for three different combinations of the magnetic field gradient, shown by the dependence of $\Delta E$ on distance (red circles), and the drive field spectrum (blue lines).
}
\label{fig:overview}
\end{figure*}

We realize programmable spin-exchange interactions in an array of atomic ensembles within an optical cavity.  Our scheme, illustrated in Fig.~\ref{fig:overview}, produces a class of spin models described by an effective Hamiltonian
\begin{equation}
H_{\text{eff}} = -\sum_{\mu,\nu} J(r_{\mu\nu}) f^+_\mu f^-_\nu + H_q. 
\label{eq:Heff}
\end{equation} Here, $f^{+(-)}_\mu$ denotes the raising (lowering) operator for the Zeeman spin of atom $\mu$, the distance between atoms is $r_{\mu\nu}$, and $H_q$ represents the quadratic Zeeman energy.  The spin-exchange coupling $J$ arises from a process in which one atom flips its spin down while scattering a photon from a drive field into the cavity and a second atom rescatters this photon to flip its spin up \cite{davis2019photon,davis2020protecting}. We focus on a system of spin-1 atoms initialized in the $m=0$ Zeeman state, where the effect of this ``flip-flop'' interaction is to produce correlated atom pairs in states $m=\pm 1$ \cite{masson2017cavity,davis2019photon,hamley2012spin}.

Whereas the single-mode cavity ordinarily mediates interactions among all sites, we break this all-to-all connectivity by introducing a magnetic field gradient along the cavity axis.  The gradient introduces a difference $\hbar\omega_B$ between the Zeeman splittings on adjacent sites, such that spin-exchange processes are off-resonant for physically separated spins. To controllably reintroduce interactions between ensembles spaced by a distance of $r$ sites, we modulate the intensity of the drive field---and hence the instantaneous spin-exchange coupling $\tilde{J}(t)$—at a frequency $r \omega_B$.  More generally, to obtain a specified set of couplings $J(r)=J^*(-r)$ in Eq. \eqref{eq:Heff}, we set a drive waveform
\begin{equation}
\tilde{J}(t) = \sum_{r} e^{-i r \omega_B t} J(r)
\end{equation}
according to the Fourier transform of the couplings.  The drive waveform thus determines the dispersion relation $\chi_k = -2n\tilde{J}(k/\omega_B)$ for spin waves with momentum $k$, in a system of $n$ atoms per site.

We probe the connectivity of interactions by measuring spatial correlations in the populations of the $m=\pm 1$ states in an array of $M=18$ sites, with $n\approx 10^4$ rubidium-87 atoms per site.  After turning on interactions for 100-200~$\micro{s}$, after which 30-50\% of the atoms are in states $m = \pm 1$, we perform state-sensitive imaging to obtain the correlations $C^\text{pm}_{ij} = \Corr{n_{+,i}}{n_{-,j}}$ in the populations $n_{+,i},n_{-,j}$ of states $m=\pm 1$ for each pair of sites $(i,j)$.  Figure~\ref{fig:overview}b shows the measured correlations for three different scenarios.  For a monochromatic drive field, in a uniform magnetic field we observe correlations of equal strength between all sites, indicating the expected all-to-all interactions.  By contrast, adding a magnetic field gradient results in correlations being localized to individual sites.  Finally, modulating the intensity of the drive light at frequency $r\omega_B$ produces correlations between all pairs of sites separated by a distance $\abs{i-j} = r$, as shown for $r=10$.

\begin{figure}[tbh]
\includegraphics[width=\columnwidth]{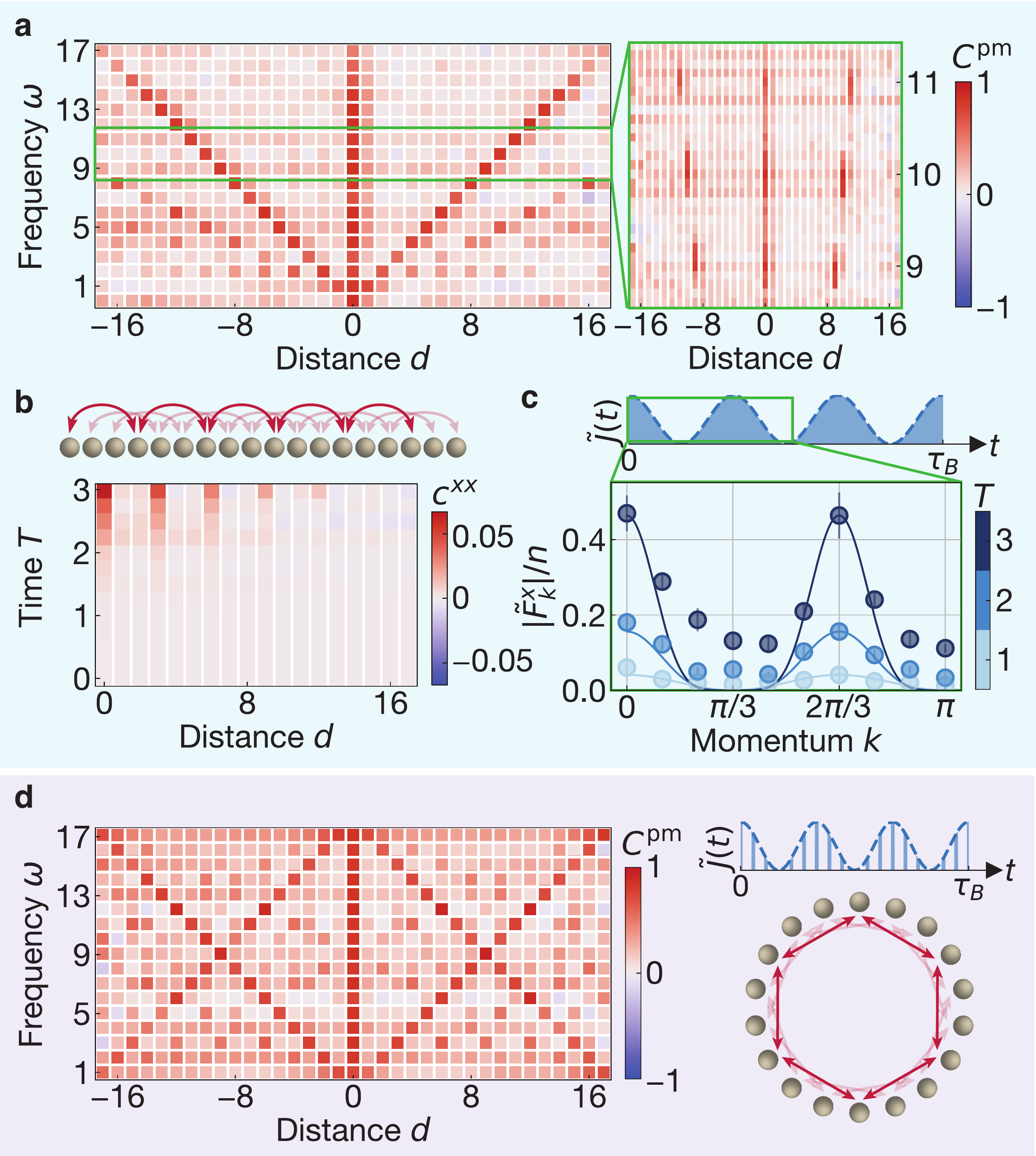}
\caption{\textbf{Pair creation at programmable distance.}  \textbf{a} Correlations $C^{\mathrm{pm}}$ vs modulation frequency $\omega$ and distance. Setting $\omega = r\omega_B$ generates correlations $C^\text{pm}$ at a distance
$r$.  Inset shows correlations $C^\text{pm}$ appearing only when satisfying a
resonance condition $\omega = r\omega_B$ for integer $r$.  \textbf{b} Measurements of $c^{xx}$ for sinusoidal modulation at frequency $3\omega_B$ reveal spreading of correlations to integer multiples of the distance $r=3$ after $T$ Bloch periods.
\textbf{c} Magnitude of the structure factor $|\tilde{F}_k^x| = \avg{\tilde{F}^x_{k}\tilde{F}^x_{-k}}^{1/2}$ after 1,2 and 3 Bloch periods, illustrating the relationship between the drive waveform $\tilde{J}(t)$ and the momentum-space dynamics. Error bars show standard error of the mean and curves show model $|\tilde{F}^x_k| \propto |\tilde{J}(k/\omega)|^T$, with amplitude as the only free parameter.
\textbf{d} Multiplying the sinusoidal modulation with a Dirac comb with frequency $M\omega_B$
generates periodic boundary conditions, evidenced by correlations at distances $r$ and $M-r$.}
\label{fig:sideband_frequency}
\end{figure}

The dependence of spatial correlations on modulation frequency is shown in Fig.~\ref{fig:sideband_frequency}a.  There, we plot the average correlation $C^\text{pm}(d)
=\sum_{i} C^\text{pm}_{i, i + d}/{(M - |d|)}$ of sites separated by distance $d$. Plotting $C^\text{pm}(d)$ as a function of modulation frequency $r\omega_B$, for integer values $r$, reveals correlations at distances $d = \pm r$ and $d=0$.  While the correlations at $d=0$ indicate on-site pair creation that is resonant even for a single drive frequency, the correlations at $d=\pm r$ confirm the presence of interactions at the distance set by the modulation frequency.  The interactions are spectrally well resolved as a function of drive frequency [Fig.~\ref{fig:sideband_frequency}a inset], highlighting the precise control of the coupling distance.

To sensitively probe the growth and spreading of correlations, we examine the transverse magnetization, which provides an enhanced signal at early times.  Specifically, we evaluate the normalized covariance ${c^{xx} = {\Cov{F^x_i}{F^x_{i + d}}}/n^2}$, where $\vec{F}_i = \sum_{\mu \in i} \vec{f}_\mu$ denotes the collective magnetization on site $i$ in a rotating frame set by the local magnetic field.  Figure~\ref{fig:sideband_frequency}b shows $c^{xx}$ as a function of time and distance $d$, averaged over all sites $i$, for a system programmed to interact at distance $r = 3$.  Correlations first appear between nearest neighbors on the coupling graph and spread over time to further neighbors at multiples of the distance $r$. We additionally compute the structure factor $\tilde{F}^x_k = \sum_l e^{i k l}F^x_l/\sqrt{M}$, plotting its rms value in Fig.~\ref{fig:sideband_frequency}c.  We observe narrowing in momentum space as function of time, complementary to the observed spreading of correlations in position space.

The growth of the structure factor is consistent with an analytical model where spin waves of momentum $k$ are amplified by a factor proportional to $\abs{\chi_k}$ per Bloch period of evolution.  Equivalently, the growth in $|\tilde{F}_k^x|$ for each momentum mode $k$ is proportional to the drive intensity $\tilde{J}(t)$ at time $t=k/\omega_B$.  The amplification is notably strongest at minima of the dispersion relation $\chi_k < 0$. Pair creation thus drives the system towards states of minimal interaction energy, while increasing the quadratic Zeeman energy $H_q$ to compensate.

Engineering the dispersion relation via the drive waveform remarkably allows for realizing periodic boundary conditions (PBC), despite the physical geometry of our array as an open chain.  For a chain of $M$ sites with PBC, the domain of the dispersion relation is a discrete set of points in momentum space, spaced by $\Delta k = 2\pi/M$.  Correspondingly, we break the drive waveform into a train of short pulses with spacing $\tau_B/M$ in time, where $\tau_B = 2\pi/\omega_B$ is the Bloch oscillation period for spin excitations. For an initial sinusoidal modulation designed to introduce interactions at distance $r$, the pulsed variant has a frequency spectrum that includes peaks at both $r\omega_B$ and $(M - r)\omega_B$.  The resulting correlations $C^\mathrm{pm}$, shown in Fig.~\ref{fig:sideband_frequency}d, are strongest at distances $d=\pm r$ and $d = \pm (M-r)$, indicating that the system now behaves as though the sites were situated on a ring.

\begin{figure*}[tbh]
\includegraphics[width=\textwidth]{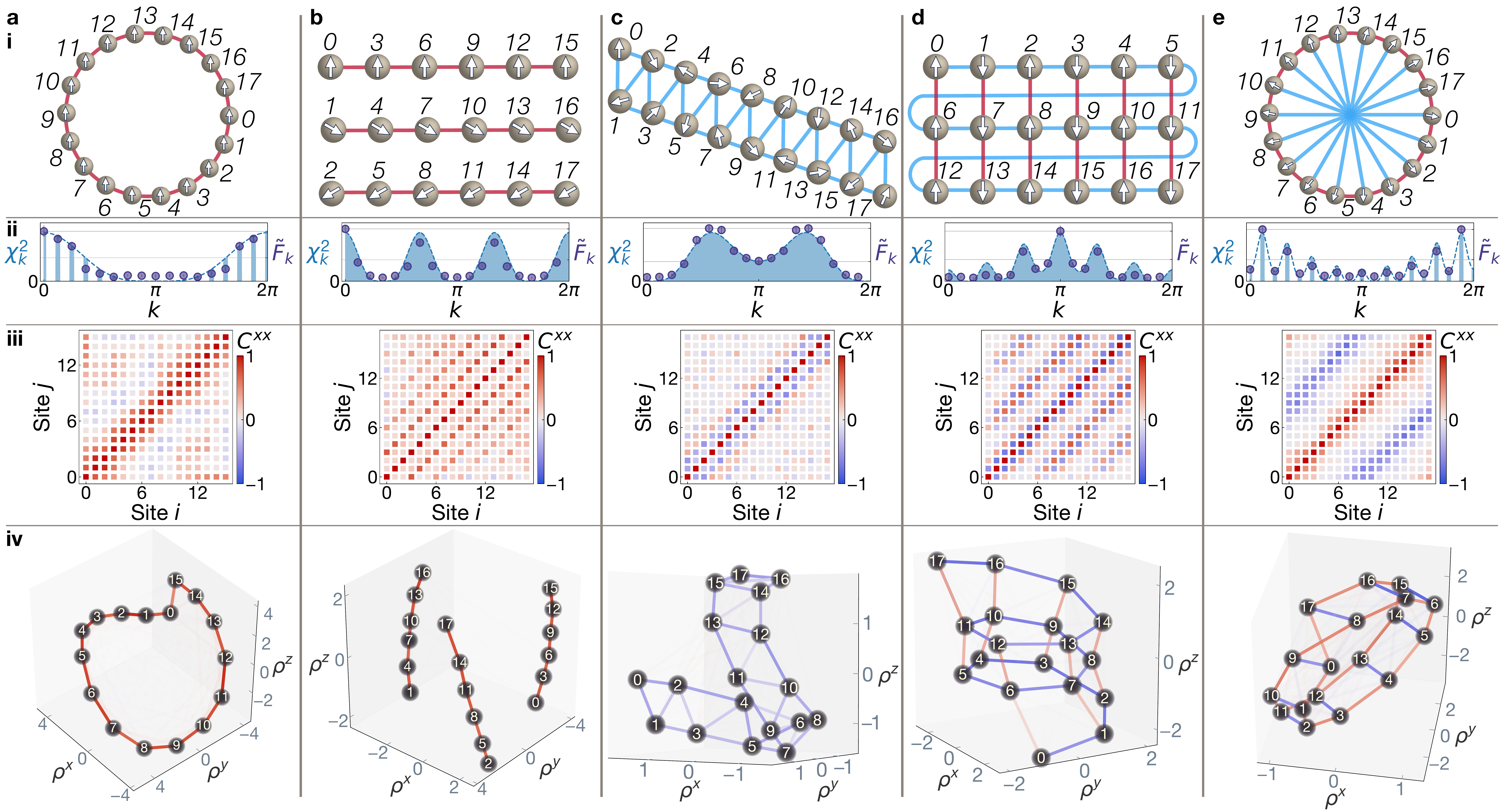}
\caption{\textbf{Geometry extracted from correlations.} 
\textbf{a} Ring, \textbf{b} disconnected chains, \textbf{c} triangular antiferromagnetic ladder, \textbf{d} cylinder with anisotropic sign of interactions, and \textbf{e} M\"obius ladder with oppositely signed interactions along the edge and width.  \textbf{i} Connectivity graphs, with red (blue) bonds indicating ferromagnetic (antiferromagnetic) couplings between sites labeled by position in array.  \textbf{ii} Structure factor (purple circles) measured after $T=2$ Bloch periods is proportional to squared dispersion relation $\chi_{k}^2$ (blue shaded region).  \textbf{iii} Correlations $C^{xx}$.  \textbf{iv} Reconstructed geometries in $D = 3$ dimensions, with Cartesian coordinates $\boldsymbol{\rho}$.  Color and opacity of bonds indicate sign and magnitude of the inferred coupling $J'$.}
\label{fig:geometry}
\end{figure*}

We verify the successful realization of periodic boundary conditions by directly reconstructing the effective geometry of the system from the measured spin correlations $\Cxx_{ij} = \Corr{F^x_i}{F^x_j}$.  Adopting an ansatz that correlations decay as a Gaussian function $\left|\Cxx_{ij}\right|\sim e^{-\abs{\boldsymbol{\rho}_i-\boldsymbol{\rho}_j}^2}$ of distance $\abs{\boldsymbol{\rho}_i-\boldsymbol{\rho}_j}$ in a $D$-dimensional space, we seek a mapping of the array sites to effective coordinates $\boldsymbol{\rho}_i$ that best fit the distances $d_{ij} = \sqrt{-\log |\Cxx_{ij}|}$ inferred from the correlations.  We obtain the coordinates $\boldsymbol{\rho}_i$ by applying metric multidimensional scaling~\cite{torgerson1952multidimensional} to the distance matrix $d_{ij}$.  The result is shown in Fig.~\ref{fig:geometry}a for a system with nearest-neighbor interactions and periodic boundary conditions. In addition to calculating the effective coordinates $\boldsymbol{\rho}_i$, we calculate an inferred coupling matrix $J' = \left(\Cxx\right)^{-1}$.  Coloring the edges between all pairs of sites according to $J'$ corroborates the ring-like coupling graph.

More broadly, tailoring the drive waveform enables versatile control over the geometry and topology of the coupling graph, as we illustrate by the same blackbox reconstruction technique.  We first observe that introducing interactions at a distance $r>1$, with open boundary conditions, produces a set of $r$ disjoint chains, as depicted in Fig.~\ref{fig:geometry}b for $r=3$.  Linking such chains with a second modulation frequency generates a two-dimensional graph, as shown by the triangular ladder in Fig.~\ref{fig:geometry}c, formed by interactions at distances $r_1=1$ and $r_2 = 2$.  Furthermore, adding periodic boundary conditions allows for realizing nontrivial topologies.  As illustrative examples, Fig.~\ref{fig:geometry}d shows a square-lattice cylinder, while Fig.~\ref{fig:geometry}e shows a M\"{o}bius ladder.  The characteristic twist of the M\"obius strip is evident in the crossing of two bonds in the reconstructed geometry.

For a given coupling graph, the sign of the interaction $J(r)$ at each distance $r$ is set by the phase of the modulation at frequency $r\omega_B$.  We always choose the on-site interaction to be ferromagnetic, favoring a large spin polarization on each site, and choose a phase $\arg[J(r)] \in\{0,\pi\}$ to set either ferromagnetic or antiferromagnetic couplings at each nonzero distance $r$.  Figure~\ref{fig:geometry} includes examples with ferromagnetic (a-b), anti-ferrogmagnetic (c) and sign-changing (d-e) couplings.

The antiferromagnetic triangular ladder in Fig.~\ref{fig:geometry}c constitutes a fully frustrated XY model \cite{lee1998phase}.  In the classical ground state of this model, adjacent spins have a relative angle of approximately $120^\circ$ [white arrows in Fig.~\ref{fig:geometry}c.i].  More precisely, the angle by which the phase winds is predicted by the peaks in $\chi_k$ at $k\approx \pm 0.58\pi$, indicating two degenerate minima in the spin-wave dispersion $\chi_k$ for two possible directions of phase winding.  The measured structure factor and spin correlations in Fig.~\ref{fig:geometry}c.ii-iii are consistent with the predicted ordering, with antiferromagnetic correlations $\Cxx < 0$ between each pair of neighboring sites on the ladder resulting in the blue bonds in Fig.~\ref{fig:geometry}c.iv. 

Our approach also allows for specifying interactions that change sign as a function of distance, as we illustrate for the cylinder and the M\"{o}bius ladder in Figs.~\ref{fig:geometry}d-e.  In each case, by choosing opposite signs of interaction for two distances $r_1, r_2$ between sites in the atom array, we obtain an anisotropic sign of the interaction on the two-dimensional manifold representing the effective geometry [Fig.~\ref{fig:geometry}d-e.iv].  For example, in the M\"{o}bius strip, the ferromagnetic interactions at distance $r_1 = 1$ give rise to ferromagnetic correlations (red bonds) all along the singular edge of the strip, while the antiferromagnetic couplings at distance $r_2=9$ are manifest in the antiferromagnetic correlations (blue bonds) across the width of the strip.  The measured correlations are indicative of the transverse magnetization winding by $2\pi$ along the closed loop formed by the edge of the strip.

Generically, engineering nonlocal couplings allows for exploring radically different geometries, beyond those that can be visualized by an embedding of sites in a Euclidean space.  Inspired by models of quantum gravity~\cite{gubser2017p,heydeman2016tensor}, we proceed to simulate a non-Archimedean geometry \cite{gubser2018continuum,bentsen2019treelike}, where the points on the real line are best viewed as leaves on an infinite regular tree graph.  Such tree graphs feature in a version of the Anti-de Sitter/Conformal Field Theory correspondence ($p$-adic AdS/CFT  \cite{gubser2017p,heydeman2016tensor}), in models of information scrambling in black holes \cite{barbon2013fast}, and in tensor-network representations of strongly correlated quantum states \cite{shi2006classical,murg2010simulating}.

\begin{figure*}[tbh]
\includegraphics[width=\textwidth]{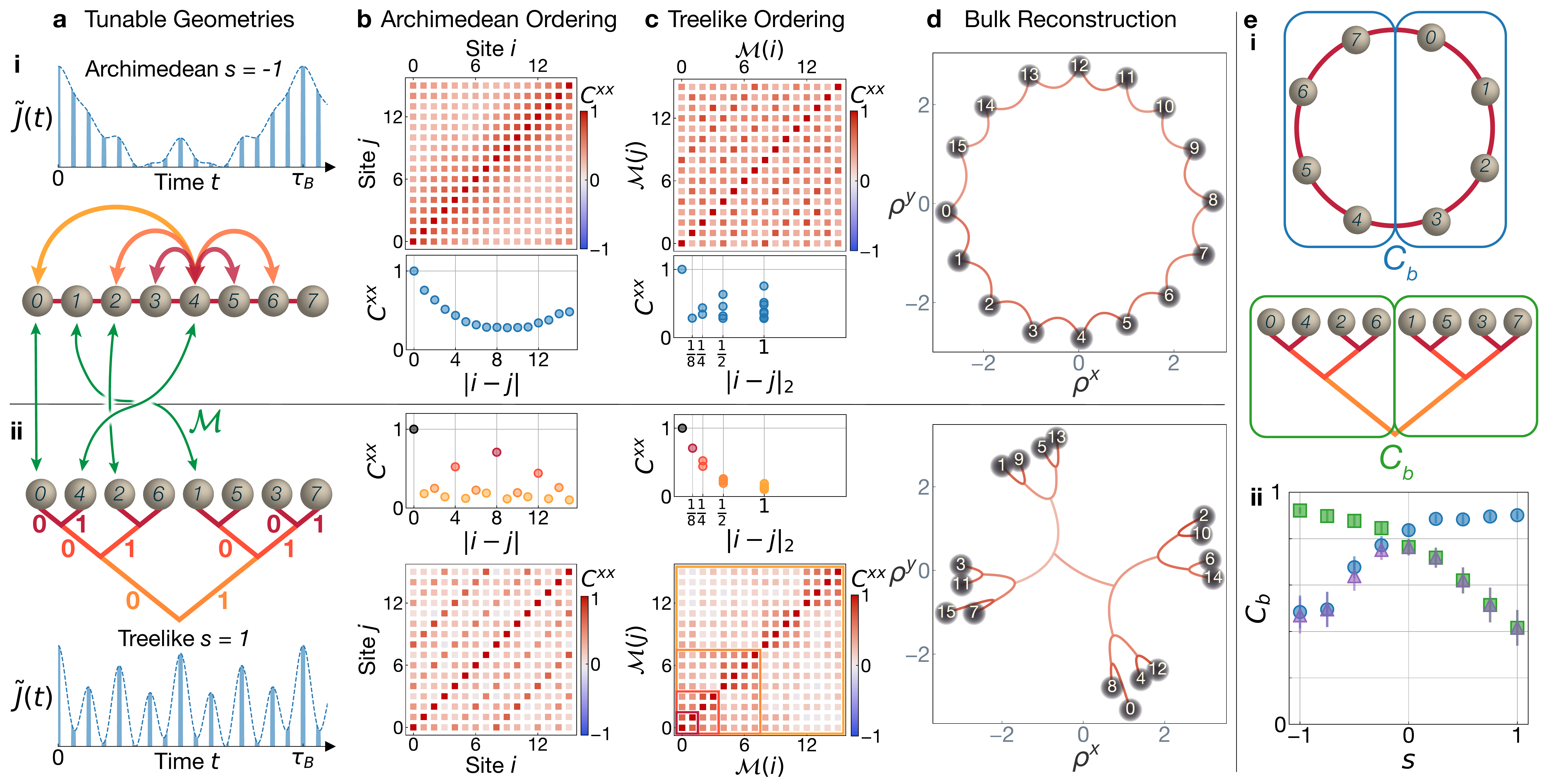}
\caption{\textbf{Treelike Geometry.} \textbf{a} Waveforms $\tilde{J}(t)$ generating \textbf{i} Archimedean and \textbf{ii} treelike interactions.  The two geometries are related via the Monna map $\mathcal{M}$ (green arrows). \textbf{b} Physical and \textbf{c} treelike orderings of $\Cxx$.  Correlations decay smoothly only as a function of Archimedean distance $\abs{i-j}$ for $s = -1$, and only as a function of treelike distance $\abs{i-j}_2$ for $s = 1$.  \textbf{d} Blackbox bulk reconstructions from $\Cxx$ reveal no bulk for $s = -1$ and a binary tree structure for $s = 1$. \textbf{e} Coarse-grained bipartite correlation $C_b$ vs. exponent $s$ for partitions according to \textbf{i} Archimedean (blue circles) and \textbf{ii} treelike (green squares) ordering of sites. The minimal correlation over all possible bipartitions (purple triangles) is peaked at $s = 0$, signifying a breakdown of locality. Error bars denote standard deviation estimated by jackknife resampling. }
\label{fig:trees}
\end{figure*}

To access a treelike geometry, we engineer couplings
\begin{equation}\label{eq:tree_couplings}
  J(i - j) \propto 
  \begin{cases}
    |i - j|^s & |i - j| = 2^n, n\in \mathbb{Z}\\
    0 & \mathrm{otherwise},
  \end{cases}
\end{equation}
where the parameter $s$ allows for tuning between Archimedean and non-Archimedean regimes \cite{bentsen2019treelike}.  For $s < 0$ the system is approximately a one-dimensional chain [Fig.~\ref{fig:trees}a.i], whereas setting $s>0$ theoretically produces the treelike geometry shown in Fig.~\ref{fig:trees}a.ii \cite{gubser2018continuum, bentsen2019treelike}.  Each leaf of the tree represents an array site, whose position in the tree is determined by branching left or right at level $\lev$ if the $\lev^\mathrm{th}$ bit of the site index $i$ is $0$ or $1$.  Starting from the base of the tree, the first branching is governed by the least significant bit, because for $s>0$ the weakest couplings are between even and odd sites.  Thus, the order of sites in the tree is rearranged from the physical order by the Monna map $\mathcal{M}(i)$, which reverses the order of bits in the site index $i$.

We confirm the transition from an Archimedean to a treelike geometry by
measurements of spin correlations $\Cxx$ for $s = \pm 1$.  We implement both models for $M=16$ sites with periodic boundary conditions, using the drive waveforms in Fig.~\ref{fig:trees}a.i-ii.  For each value of $s$, we show $\Cxx$ as a function of physical site indices $i,j$ [Fig.~\ref{fig:trees}b] and as a function of positions $\mathcal{M}(i), \mathcal{M}(j)$ on the tree [Fig.~\ref{fig:trees}c].  Whereas for $s = -1$ we observe a smooth decay of correlations as a function of physical distance, for $s = 1$ we observe a non-monotonic dependence of correlations on physical distance due to the highly nonlocal structure of interactions.  The Monna-mapped correlations for $s=1$, however, are strongest near the diagonal---indicating a new sense of locality in the non-Archimedean geometry---and exhibit blocks consistent with the hierarchical structure of the tree.

To corroborate the realization of a non-Archimedean geometry, we plot the dependence of correlations on a treelike measure of distance in Fig.~\ref{fig:trees}c.  The natural metric for the treelike geometry is the 2-adic norm $|r|_2 = 2^{-\lev}$, where $\lev$ is the largest integer such that $r$ is divisible by $2^\lev$.  Intuitively, the $2$-adic distance $|i-j|_2 = 2^{-\lev}$ between sites $i$ and $j$ is governed by the level $\lev$ of the tree --- counting up from the base --- at which the leaves representing the two sites connect.  As a function of $2$-adic distance, we observe a smooth decay of correlations.

A key feature of the tree graph is that only the vertices on the boundary represent physical sites, whereas the interior vertices constitute a holographic bulk geometry embodying the effective distance between sites.  To investigate the validity of this holographic description, we perform a blackbox reconstruction of the bulk geometry from spin correlations.  We begin by mapping the physical sites to effective coordinates in a Euclidean space, as before.  Next, we draw bonds between pairs of maximally correlated sites, representing permutations between sites that minimally disturb the unknown bulk geometry. Specifically, we find the distance $r$ that maximizes the correlations $\Cxx(r)$ and connect all sites separated by a distance $\abs{i-j}=r\bmod M$. We then adopt a coarse-graining procedure, treating each pair or group of connected sites as a new larger site, and drawing new connections, until there is a path through the bulk between any two sites on the boundary.  

The bulk reconstructions are shown in Fig.~\ref{fig:trees}d for both the Archimedean ($s=-1$) and non-Archimedean ($s=1$) cases.  For $s = -1$, where interactions between physical neighbors dominate, the reconstruction produces only a one-dimensional loop.  By contrast, for $s = 1$, a tree emerges from the reconstruction as a bulk geometry encapsulating the structure of spin correlations. This emergent geometry is analogous to the gravitational bulk in the $p$-adic AdS/CFT correspondence, where the tree serves as a discretized version of hyperbolic space \cite{gubser2017p,heydeman2016tensor}.

The transition between two radically different geometries depending on the sign of the exponent $s$ suggests that all sense of locality is lost as $s$ approaches zero.  To probe the breakdown of locality, we consider different possible bipartitions of the $M=16$ sites into 8-site subsystems $I$ and $J$ and examine correlations between the subsystems [Fig.~\ref{fig:trees}e].  Specifically, we plot a bipartite correlation $C_b = \Corr{F^x_I}{F^x_J}$, where $F^x_I = \sum_{i\in I} F^x_i$ denotes a coarse-grained spin, as a function of $s$.  For $s<0$, the correlation $C_b$ is smaller for a cut that is local according to the physical ordering of sites (blue circles) than for a cut that is local on the tree (green squares), whereas for $s>0$ the situation is reversed, consistent with the change in effective geometry.  Further plotting the minimum correlation $C_b$ over all possible bipartitions (purple triangles) reveals a peak at $s=0$, indicating the absence of any geometry providing a sense of locality.

The breakdown of locality at $s=0$ paves the way towards studies of fast scrambling \cite{bentsen2019treelike}, the generation of system-wide entanglement at a conjectured maximal possible rate --- that of a black hole~\cite{hayden2007black,sekino2008fast}.  More broadly, our work provides a starting point for harnessing quantum simulators to investigate the conjecture that spacetime geometry and gravity are emergent phenomena arising from entanglement among microscopic degrees of freedom~\cite{qi2018does}.  The treelike geometry can serve as a model for probing transport through the holographic bulk and enable the implementation of holographic error-correcting codes~\cite{Pastawski2015, heydeman2016tensor}.  Further, our reconstruction of the bulk offers a blueprint for searching for gravitational duals in a wide range of quantum many-body systems.

The antiferromagnetic and sign-changing interactions demonstrated here open new opportunities for studies of frustrated magnetism.  Introducing disorder will allow for realizing spin-glass models~\cite{gopalakrishnan2011frustration,strack2011dicke} that map to NP hard problems in pattern recognition~\cite{Amit1985,marsh2021enhancing} and optimization~\cite{berloff2017realizing}.  The dynamics of pair creation might be harnessed to find ground states by gain-based optimization~\cite{mcmahon2016fully,berloff2017realizing} and to investigate the computational benefit of entanglement.  Our scheme also generalizes to implementing synthetic gauge fields by introducing complex-valued couplings \cite{rudner2020band}, for explorations of topological physics.

Programmable pair creation can further be leveraged to engineer entangled states applicable to sensing \cite{hamley2012spin,masson2017cavity} and computation.  Control over the spatial structure of entanglement will enable enhanced sensing and imaging of spatially extended fields~\cite{Pezze2018}.  Our method also enables the generation of continuous-variable graph states, a resource for measurement-based quantum computation, as well as tensor-network states~\cite{shi2006classical,murg2010simulating} applicable to hybrid quantum-classical algorithms.  While our experiments benefit from collectively enhanced interactions among ensembles, a regime with a single quantum spin per site could be accessed with Rydberg-blockaded ensembles, with individually trapped atoms in a cavity or waveguide deep in the strong-coupling regime \cite{hung2016quantum}, and in extensions to trapped ions \cite{ozeri2020quantum} or color centers \cite{evans2018photon}.

\section*{Methods}

\subsection{Experimental Sequence}
We begin by loading rubidium-87 atoms from a magneto-optical trap (MOT) into an array of microtraps, where we use optical pumping and adiabatic microwave sweeps to prepare the atoms in the $\ket{F = 1, m = 0}$ state. We then transfer the atoms into a 1560\,nm optical lattice supported by the cavity, resulting in a set of $M = 16$ or 18 discrete ensembles.
To generate programmable interactions between the ensembles, we apply a magnetic field gradient and drive the optical cavity along its axis with a modulated intensity. 
After the interaction time, we load the atoms back into the microtraps and use state-selective fluorescence imaging to measure the population in each Zeeman state.  To measure the transverse magnetization, we apply a series of local spin rotations prior to the imaging sequence.  Ext. Data Fig.~\ref{fig:sequence}a shows a schematic of the experimental sequence.

\subsection{Microtraps and Lattice Transfer}
Our experiments employ a hybrid trapping scheme: whereas we perform cooling, internal state preparation, and imaging in a microtrap array, we transfer the atoms to an intracavity optical lattice before inducing cavity-mediated interactions.  The 1560\,nm intracavity lattice is in registry with the standing wave of 780\,nm light used to drive interactions, and thus maximizes the atom-light coupling.  However, because the 1560\,nm light produces a strong and inhomogeneous ac Stark shift of the $5P_{3/2}$ state \cite{lee2014many}, we instead use the 808\,nm microtrap array during portions of the experimental sequence requiring near-resonant light, namely cooling, optical pumping, and fluorescence imaging.

We initially turn on a two-dimensional array of $M\times 2$ optical microtraps at 808\,nm during MOT loading.  The long axis of the array is aligned with the cavity axis, with $60\,\micro{m}$ between traps.  The two transverse traps are designed to double the total trap volume and, correspondingly, the number of atoms loaded into the cavity for a fixed microtrap waist.  Each microtrap has a waist of $6\,\micro{m}$ and a depth of $h\times 4\,\text{MHz}$.   During the loading phase the transverse microtrap spacing is $100\,\micro{m}$.  After loading the microtraps, the transverse spacing is reduced to $8\,\micro{m}$, so that both transverse traps fit within the $25\,\micro{m}$ waist of the intracavity lattice. We adiabatically transfer the atoms from the microtraps into the intracavity lattice, increasing the lattice power from an initial depth of $h\times 200\,\text{kHz}$ to $h\times 3.5\,\text{MHz}$ and then ramping off the microtraps. This preparation results in 1D array of $M$ ensembles at a temperature of $100\,\micro{K}$, with each ensemble containing $n \approx 10^4$ atoms spread over 10 lattice sites. 

For imaging, we transfer the atoms from the optical lattice back into the microtrap array by first switching on the $M\times 2$ microtraps before reducing the lattice depth to $h\times 200\,$kHz. Subsequently, we adiabatically move the microtraps away from the optical lattice by approximately 15\,$\mu$m to avoid ac Stark shifts during imaging.

\subsection{Imaging and Spin Readout}
We detect the atoms in a sequence of four fluorescence images designed to independently measure the populations of all three Zeeman states within the $F=1$ manifold and any residual atoms in $F=2$. 
For each fluorescence image we apply a retro-reflected laser beam resonant with the $F=2 \rightarrow F' = 3$ transition of the D2 line for 100\,$\micro{s}$ and collect the resulting fluorescence signal on an EMCCD camera.
With the first imaging pulse, we measure the population in the $F=2$ manifold, expelling these atoms from the microtraps by heating.
For state-selective imaging of the $F = 1$ manifold, we sequentially apply three microwave sweeps which adiabatically transfer the atoms from each magnetic substate to $F=2$ and perform fluorescence imaging after each sweep.
A typical fluorescence signal of the atoms is shown in Ext. Data Fig.~\ref{fig:sequence}b.
For background subtraction we use a method from Xu et al.~\cite{xu2019probing} based on a principal component analysis of approximately 100 images without atoms.

To measure the transverse spin component $F^x_i$ we sequentially perform local spin rotations at each site $i$ prior to the imaging sequence. For this purpose, we focus a circularly polarized laser onto each site by controlling the position of the beam with an acousto-optic deflector. By modulating the intensity of the laser at the local Larmor frequency, we induce a resonant Raman coupling between adjacent magnetic sublevels. We apply a $3\,\mu$s Raman pulse to produce a $\pi/2$ spin rotation. This locally maps $F^x_i$ onto the measurable population difference $n_{+,i} - n_{-,i}$, illustrated in Ext. Data Fig.~\ref{fig:sequence}c. Here, $F^x_i$ is defined in a rotating frame that depends on the local Larmor frequency at site $i$.
Shot-to-shot fluctuations in the Larmor frequency lead to a reduction of measurable correlations between two sites, where the reduction depends on the time between the corresponding Raman pulses. Thus, to suppress any bias in the measured correlations, we randomize the order of the local spin rotations in each experimental realization.

\subsection{Computation of Correlations}

When visualizing the distance-dependence of interactions, reconstructing effective geometries, or probing bipartite correlations, we compute correlation functions $C^\mathrm{pm}$,  $\Cxx$, and $C_b$ from a minimum of 50 measurements.  Each correlation function is defined in the main text in terms of specified observables $A$ and $B$ as
\begin{equation}
    \Corr{A}{B} = \frac{\Cov{A}{B}}{\sqrt{\Var{A}\Var{B}}},
\end{equation}
where $\Cov{A}{B} \equiv \avg{AB}-\avg{A}\avg{B}$ and $\Var{A} \equiv \Cov{A}{A}$.  These correlations are normalized to the shot-to-shot variance, which provides the relevant spatial information while being agnostic to the the total amount of pair creation. Effects of finite statistics on the measured correlations $C^\mathrm{pm}$ are examined in Ext. Data Fig.~\ref{ExtFig:FiniteStat}.

To quantify pair creation dynamics, we measure in the $\hat{x}$-basis and normalize the covariance matrices to the population of atoms on each site rather than their variance,
\begin{equation}
    c^{xx}_{ij} = \frac{\Cov{F_l^x}{F_m^x}}{n^2} \approx \frac{\langle F_l^x F_m^x \rangle}{n^2}.
\end{equation}
For this correlator, the measurement in the $\hat{x}$-basis provides a high sensitivity at early times and a large dynamic range for measurements over time.
The normalization is chosen such that  the extracted correlation is sensitive to the total amount of pair creation, allowing us to visualize the growth of correlations as a function of time.

\subsection{Interaction Parameters}
To enable the programmable interactions, we apply a magnetic field gradient parameterized by the difference $\omega_B$ in Zeeman splittings between adjacent array sites.  This gradient is superposed on an overall bias field $B_0$ perpendicular to the cavity axis, which produces a Zeeman splitting of $\omega_z/B_0 = 2\pi\times700\,\text{kHz}/\text{G}$ and a quadratic Zeeman shift of $q/B_0^2 = 2\pi\times 72\,\text{Hz}/\text{G}^2$. We work in a regime where $\omega_z/M > \omega_B > q$, i.e., the variation in the magnetic field is small compared to the average field yet results in a Bloch oscillation frequency larger than the quadratic Zeeman shift. Specifically, we choose a magnetic field between 2 and 4\,G (as detailed in Ext. Data Table~\ref{ExtTab:Parameters} for each data set) and a typical gradient $\omega_B = 2\pi\times 1.52(1)\,\text{kHz/site}$. For measurements of $C^{\text{pm}}$ in Figs.~\ref{fig:overview}-\ref{fig:sideband_frequency}, we increase the ratio $\omega_B/q$.  This is accomplished either by increasing $\omega_B$ to $2\pi\times 12.47(2)\,\text{kHz}$, or by reducing the effective quadratic Zeeman shift to $q = 2\pi\times 60\,\text{Hz}$ by applying an ac Stark shift to the $\ket{1,0}$ state via off-resonant microwave coupling to $\ket{2,0}$.


We induce spin-exchange interactions among the atoms by applying a drive field which typically has a detuning between $\delta_c = -2\pi\times 4\,\text{MHz}$ and $-2\pi\times7\,$MHz from cavity resonance. The cavity mode itself has a large detuning of $\Delta = -2\pi\times 11\,$GHz from atomic resonance. The drive field is linearly polarized at an angle of 55 degrees with respect to the magnetic field, chosen to eliminate tensor light shifts. The instantaneous spin-exchange coupling is given by $-\tilde{J} \approx \bar{n}_\text{ph}\Omega^2/(2\delta_c)$, where $\Omega = 2\pi \times 13\,\text{Hz}$ is the vector ac Stark shift per circularly polarized photon in the cavity.  Our typical peak intracavity photon number $\bar{n}_\text{ph} = 10^4$ corresponds to a collective interaction strength $2n\tilde{J} = 2\pi\times 3\,\text{kHz}$ between ensembles of $n=10^4$ atoms.

To produce a set of couplings $J(r)$, we modulate the intensity of the drive field via an acousto-optic modulator as 
\begin{equation}
\tilde{J}(t)  =  2\sum_{r>0} \left[\cos(r \omega_B t+\phi_r)+1\right]\, |J(r)|
\end{equation}
where we use the phases $\phi_r \in \{0,\pi\}$ to set the sign of the interactions. The coupling at $r = 0$ is given by $J(0) = 2\sum_{r>0} |J(r)|$.
To produce periodic boundary conditions in the system of $M$ sites, we additionally pulse the drive at a frequency of $M\omega_B$. Each pulse has a duration of $0.3 \tau_B/M = 11\micro{s}$.

\subsection{Cavity Parameters}

The atoms are coupled to a near-concentric Fabry-Perot cavity with a length of 5\, cm and an $18\,\micro{m}$ waist at 780\,nm.  The cavity has vacuum Rabi frequency of $2g = 2\pi \times 2.6 \,\text{MHz}$ and linewidth $\kappa = 2\pi\times 250(20)\,\text{kHz}$, yielding a single-atom cooperativity $\eta = \frac{4g^2}{\kappa\Gamma} = 4.5$, where $\Gamma = 2\pi \times 6.07 \,\text{MHz}$ is the linewidth of the $5P_{3/2}$ state in rubidium.  Our drive field is detuned by $\Delta =- 2\pi \times 11 \,\text{GHz}$ from the $\ket{5S_{1/2}, F=1} \rightarrow \ket{5P_{3/2}}$ transition, which produces a vector light shift per circularly polarized photon of $\Omega_0 = -\frac{g^2}{6\Delta} = 2\pi\times26\,\text{Hz}$ on a maximally coupled atom at cavity center. For an average atom this dispersive coupling is reduced to $\Omega = 2\pi\times 13\,\text{Hz}$, primarily by thermal motion. The Rayleigh range of the cavity is $z_R = 1.3\,\text{mm}$, and each ensemble is within $0.4 z_R= 520\,\micro{m}$ of cavity center.  Displacement from cavity center contributes up to a 20\% reduction in coupling for the most distant atoms.

\subsection{Interaction Hamiltonian}


In Eq.~\eqref{eq:Heff}, we describe the distance-dependent spin-exchange interactions by a static effective Hamiltonian $H_I$, with the spin on each site defined in a rotating frame set by the local magnetic field.  Here we summarize the derivation of the effective Hamiltonian starting from the full time-dependent Hamiltonian $H_\mathrm{lab}$ in the lab frame.  The Hamiltonian $H_\mathrm{lab}$ for the spin system, obtained by adiabatically eliminating the cavity mode \cite{davis2019photon,davis2020protecting}, is given by \begin{equation}\label{eq:Hlab}
H_\text{lab}=-\tilde{J}(t) \sum_{l,m} F^+_l F^-_m + \sum_l h_l F^z_l + H_q,
\end{equation}
in terms of the collective spin $\vec{F}_l = \sum_{\mu \in l} \vec{f}_\mu$ on each site $l$, the local magnetic fields $h_l = \omega_B l$, and the quadratic Zeeman shift $H_q = q \sum_\mu (f^z_\mu)^2$, in units where $\hbar = 1$. Moving to a rotating frame with $H_0 = \sum_l h_l F^z_l$, the Hamiltonian becomes 
\begin{equation}
  H(t)=-\tilde{J}(t) \sum_{l,m} e^{i(l-m)\omega_B t} F^+_l F^-_m + H_q.
  \label{eq:RotatingFrameFull}
\end{equation}

When the collective interaction strength and quadratic Zeeman shift are weak compared to the gradient ($nJ, q \ll \omega_B$), the effective Hamiltonian is given to first order by the time average of Eq.~\eqref{eq:RotatingFrameFull}. The interaction component of the resulting effective Hamiltonian $H_I + H_q$ is
\begin{equation}
  H_I = -\sum_{l,m} J(l-m) F^+_l  F^-_{m},
\end{equation}
where 
\begin{equation}
   J(r) = \frac{1}{T} \int_0^T dt\, e^{i r \omega_B t} \tilde{J}(t).
\end{equation}
The dependence $J(r)$ of the couplings on distance is thus given by the Fourier transform of the drive waveform.

\subsection{Momentum-Space Dynamics}

To analytically compute the dynamics of the system, we write the Hamiltonian without approximation in terms of spin-wave operators $\tilde{F}^+_{k=\omega_B t} \equiv \frac{1}{\sqrt{M}}\sum_{l} e^{-ikl} {F}^+_l$, as
\begin{equation}
H(t) = -M \tilde{J}(t) \tilde{F}^+_{-\omega_B t} \tilde{F}^-_{\omega_B t} + H_q.
\end{equation}
We can understand this Hamiltonian by recognizing that, in the lab frame, the magnetic field gradient causes spin waves to undergo Bloch oscillations at frequency $\omega_B$. Only spin waves with momentum $k=0$ in the lab frame couple to the cavity. In the rotating frame set by the gradient, the same physics can be viewed as spin waves remaining static over time while the mode to which the cavity couples is given by $k=\omega t$. The quadratic Zeeman shift is left unchanged by the change of reference frames.

Since the system is finite and discrete, there are only $M$ orthogonal momentum modes. To obtain a discrete set of momentum-space couplings $\tilde{J}_k = -\chi_k/2n$, we drive interactions with a pulsed drive $\tilde{J}(t) = \sum_k \frac{2\pi}{M} \tilde{J}_k \delta(\omega_Bt-k)$ that only takes on non-zero values $M$ times per Bloch period. We observe that the momentum modes decouple in the Hamiltonian,
\begin{equation}
H(t) = -\sum_k 2\pi \tilde{J}_k \delta(t\omega_B-k) \tilde{F}^+_{-k} \tilde{F}^-_{k} + H_q.
\label{eq:PulsedH}
\end{equation}
The evolution of any given momentum mode is discrete, with a short period of coupling to the optical cavity that induces spin-spin interactions, followed by a longer period of time when the state evolves only under the quadratic Zeeman shift. In the limit of a large collective interaction strength $|\chi_k| > \omega_B$, each momentum mode grows by a factor of $\lambda \approx |2\chi_k \tau_B| \sin(q\tau_B)$ after each Bloch period (see Supplementary Information). This growth is reflected by the structure factor, with ${|\tilde{F}^x_{k}| \propto |\chi_k|^T \propto |\tilde{J}(k/\omega_B)|^T}$ after $T$ Bloch periods.

While our derivation of the growth of the structure factor assumes a pulsed drive field, which produces periodic boundary conditions, the same relation provides a good approximation in the case of a continuous drive field that produces open boundary conditions.  In the latter case, we expect small deviations from the model because the cavity couples to a continuum of non-orthogonal momentum modes.  We compare the continuous and pulsed cases in a numerical simulation presented in Ext. Data Fig.~\ref{ExtFig:Comp_Sim_Data}.

A key feature of the evolution in momentum space is that the modes with minimum energy are maximally amplified in our system with $\chi_k<0$.  We can gain additional intuition for this effect by considering the limit where the dynamics are slow compared to the Bloch period and a time-averaged Hamiltonian is valid. In this case, the dynamics for each momentum mode are identical to the single-mode case that has been studied previously \cite{davis2019photon,stamper2013spinor}. The system is unstable to pair creation when the collective interaction strength $2\chi_k = -4n\tilde{J}_k$ has a greater magnitude and opposite sign from the quadratic Zeeman shift $q$. This condition motivates our choice of ferromagnetic on-site interactions, such that ${\chi_k < 0}$, in our system with $q>0$. The opposite signs of $\chi$ and $q$ allow the system to access low-energy states of the interaction Hamiltonian $H_I$ by transferring energy into $H_q$ via pair creation. 

\subsection{Euclidean Reconstruction}
We leverage our understanding of the pair-creation dynamics to reconstruct effective coordinates $\boldsymbol{\rho}$ and inferred couplings $J'$ directly from measured correlations $\Cxx$.  Specifically, building on our analytical model for the growth of the structure factor, we here derive the Gaussian ansatz for the decay of correlations with distance in the effective geometry. The dynamical evolution produces low energy states of the XY Hamiltonian, which additionally allows us to relate the inverse correlation matrix and the inferred couplings.

To analytically motivate the Gaussian ansatz used for reconstructing effective geometries, we begin by relating the structure factor to the correlations we measure in the $\hat{x}$-basis,
\begin{equation}
\begin{aligned}\label{eq:WK}
\Cxx_{lm}\sim\langle F_l^x F_m^x \rangle &= \frac{1}{M} \sum_{k_1, k_2} e^{i (k_1 l -  k_2 m)} \langle \tilde{F}_{k_1}^x \tilde{F}_{-k_2}^x \rangle \\ &= \frac{1}{M} \sum_k e^{ik(l-m)} |\tilde{F}^x_{k}|^2.
\end{aligned}
\end{equation}
The final equality holds when the momentum modes are independent from one another, such that cross terms with $k_1 \neq k_2$ go to zero. This is true either when periodic boundary conditions are imposed or in the limit of an infinite system.  Equation~\eqref{eq:WK} allows for predicting the form of spatial correlations from the dispersion relation $\chi_k$, which governs the growth of the structure factor.

As an illustrative example, we consider nearest-neighbor interactions created by the drive waveform $\tilde{J}(t) \propto (\cos \omega_B t + 1)$, corresponding to the dispersion relation
\begin{equation}
    \chi_k \propto \left(e^{ik/2} + e^{-ik/2}\right)^2.
\end{equation}
Since the correlations are the Fourier transform of the squared magnitude of the structure factor, we write an expansion of $|\tilde{F}^x_k|^2$ in terms of powers of $e^{ik}$. Recalling $|\tilde{F}^x_k|\propto |\chi_k|^T$ after $T$ Bloch periods of evolution, we compute
\begin{equation}
\begin{aligned}
    |\tilde{F}^x_k|^2 &\propto \left(e^{ik/2} + e^{-ik/2}\right)^{4T} \\
    &=\sum_{d = -2T}^{2T} {{4T} \choose {d+2T}} e^{-ikd}.
\end{aligned}
\end{equation}
The coefficients in this expansion are Fourier components corresponding to correlations at distance $d$. Thus, we have $C^{xx}(d) \propto {{4T} \choose {d+2T}}$. This binomial coefficient tends to a Gaussian function of distance $d$ after several Bloch periods, analogously to a diffusion process.

More generally a multi-frequency drive leads to diffusion within the effective geometry set by the couplings. For a generic drive that produces a dispersion relation $\chi_k \propto -\sum_r J(r)e^{ikr}$, correlations in position space are given by terms in the multinomial expansion of $|\chi_k|^{2T}$. When $J(r)>0$ this directly corresponds to a random walk within the effective geometry set by the couplings $J(r)$. Motivated by the exact result for spreading in 1D, we use a Gaussian ansatz for the correlation matrix to infer distances and hence the coordinates $\boldsymbol{\rho}$ within the effective geometry.


We motivate the inferred coupling matrix $J' = \left(\Cxx\right)^{-1}$ by recalling that a population growth rate given by $|\chi_k|$ generates a low energy state of the XY model, $H_I$.  We approximate the final state as thermal, with large inverse temperature $\beta$. We make use of the $SO(2)$ symmetry to note that $\avg{F^x_i F^x_j}$ and $\avg{F^y_i F^y_j}$ are equivalent.  Now, to compute $\avg{F^x_i F^x_j}$, we integrate over phase space, with a Boltzmann weighting $\exp(-\beta H_I)$.  To constrain the overall spin length, we introduce a chemical potential $\mu$, so that
\begin{equation}
\begin{aligned}
  \Cxx_{ij}\propto \prod_l \left[\int dF^x_l\right]\, F^x_i F^x_j\exp\left(-\beta \left[H_I - \mu(F^x_l)^2\right]\right).
\end{aligned}
\label{eq:gaussian_moment}
\end{equation}
The chemical potential can be incorporated into a modified coupling matrix ${J_{ij}' = J_{ij} -\mu\delta_{ij}}$.  Evaluating the integral yields $\Cxx\propto J_{ij}'^{-1}$.  For the purposes of the reconstruction in Fig.~\ref{fig:geometry}, where we color bonds between sites according to $J_{ij}'$, only the off-diagonal terms of $J_{ij}'$ are relevant.

The inverse correlation matrix, also known as the concentration or precision matrix, can also be interpreted as the partial correlation matrix~\cite{lauritzen1996graphical}, up to normalization.  For a given set of variables $x_i$, the partial correlation between $x_i$ and $x_j$ is the correlation after regressing out every $x_{l\neq i, j}$.  In a system with interactions at distance $r$, sites spaced by $r$ have a non-zero partial correlation, but sites at distances that are multiples of $r$ have zero partial correlation, since the interactions between the sites at distance $r$ mediate all the variance.  Thus, the interpretation of the inverse correlation matrix as an inferred coupling matrix ${J' \propto \left(\Cxx\right)^{-1}}$ is well motivated even at early times, when correlations are still spreading across the system.
\begin{acknowledgments}
\input{AuxTexFiles/end_matter}
\end{acknowledgments}
\normalem
\putbib[programmable]
\end{bibunit}
\cleardoublepage

\onecolumngrid
\setcounter{table}{0}
\renewcommand{\tablename}{\textbf{Extended Data Table}}
\renewcommand{\thetable}{\textbf{\arabic{table}}}
\setcounter{figure}{0}
\renewcommand{\figurename}{\textbf{Extended Data Fig.}}
\renewcommand{\thefigure}{\textbf{\arabic{figure}}}

\section*{Extended Data}
\begin{table}[h]
    \begin{ruledtabular}
    \begin{tabular}{l|c|c|c|c}
        Data set & Magnetic field $B_0$ [G] &  Gradient $\omega_B$ [kHz/site] & Quadratic Zeeman shift $q$ [Hz] & Interaction time $T\tau_B$ [ms] \\
        \hline
        Fig.~\ref{fig:overview}b all-to-all & 2.8 & 0 & $2\pi\times 580$ & 0.1 \\
        Fig.~\ref{fig:overview}b localized & 3.8 & $2\pi\times 12.46$ & $2\pi\times 1100$ & 0.2 \\
        Fig.~\ref{fig:overview}b distance & 3.8 & $2\pi\times 12.46$ & $2\pi\times 1100$ & 0.2 \\
        Fig.~\ref{fig:sideband_frequency}a & 3.8 & $2\pi\times 12.46$ & $2\pi\times 1100$ & 0.2\\
        Fig.~\ref{fig:sideband_frequency}bc & 2.0 & $2\pi\times 1.53$ & $2\pi\times 290$& up to 1.97\\
        Fig.~\ref{fig:sideband_frequency}d & 2.0 & $2\pi\times 1.52$ & $2\pi\times 70$& 3.95\\
        Fig.~\ref{fig:geometry} & 2.0 & $2\pi\times 1.52$ & $2\pi\times 290$ & 1.32 \\
        Fig.~\ref{fig:trees} & 2.0 & $2\pi\times 1.52$ & $2\pi\times 290$ & 1.32
    \end{tabular}
    \end{ruledtabular}
    \caption{\textbf{Experimental parameters.} Magnetic offset field $B_0$, Bloch oscillation frequency $\omega_B = 2\pi/\tau_B$, quadratic Zeeman shift $q$, and interaction time $T\tau_B$ for each of the data sets presented in Figs. \ref{fig:overview}-\ref{fig:trees}.}
    \label{ExtTab:Parameters}
\end{table}
\newpage
\begin{figure}[htb]
    \includegraphics[width = 100mm]{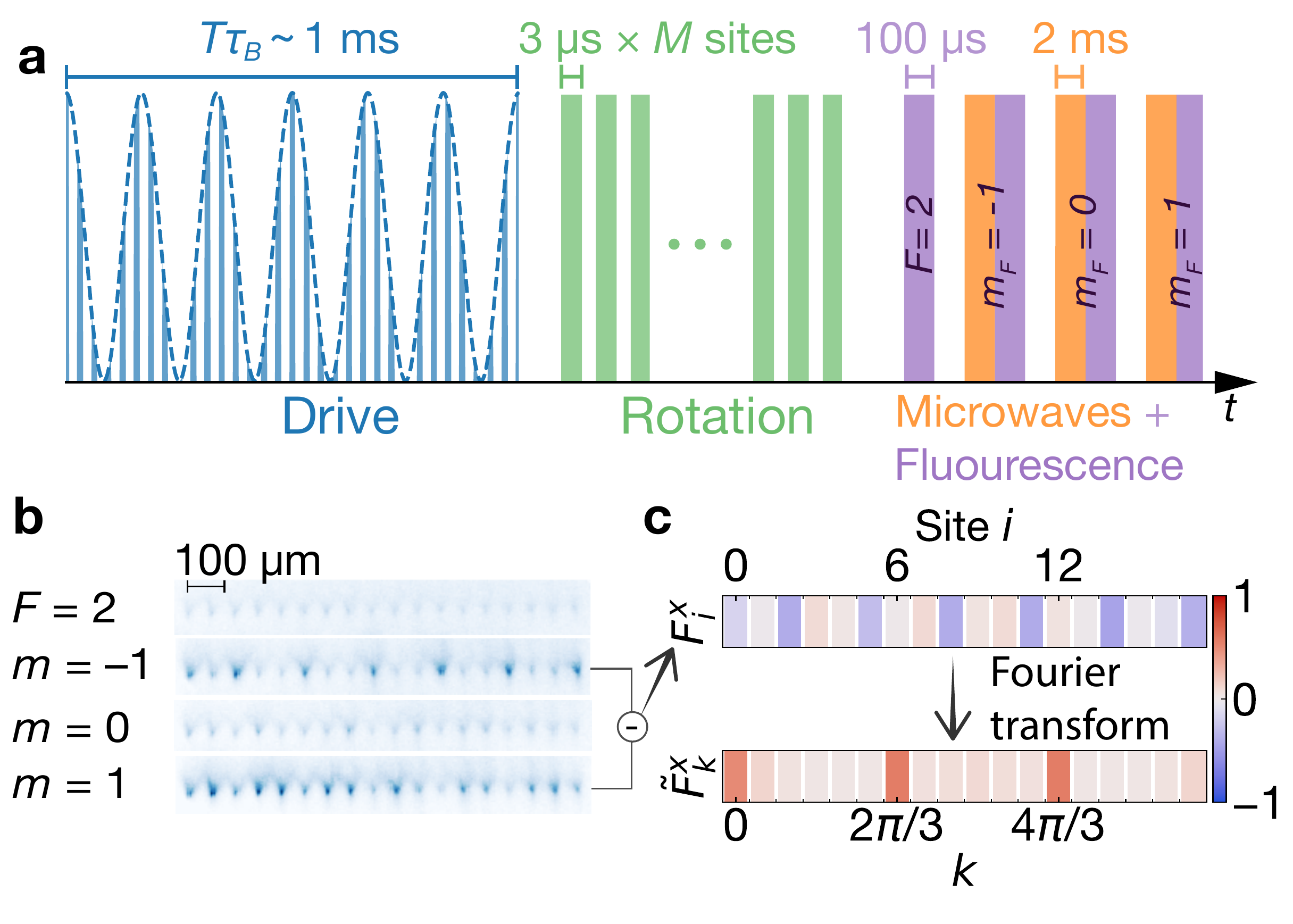}
    \caption{\textbf{Experimental sequence and imaging.} \textbf{a} Schematic of experimental sequence for measurements of $F^x_i$.  After driving the cavity to induce interactions, we apply spin rotations sequentially to the $M$ sites of the array and subsequently perform state-sensitive readout via fluorescence imaging.
    \textbf{b} Fluorescence images after spin rotation, showing the signal for the $F=2$ manifold and the three magnetic substates for the case of interactions at distance $r = 3$ with periodic boundary conditions. \textbf{c} Transverse magnetization $F^x_i$ and structure factor $\tilde{F}^x_k$ extracted from the image in \textbf{b}.}
    \label{fig:sequence}
\end{figure}
\newpage
\begin{figure}[h]
    \centering
    \includegraphics{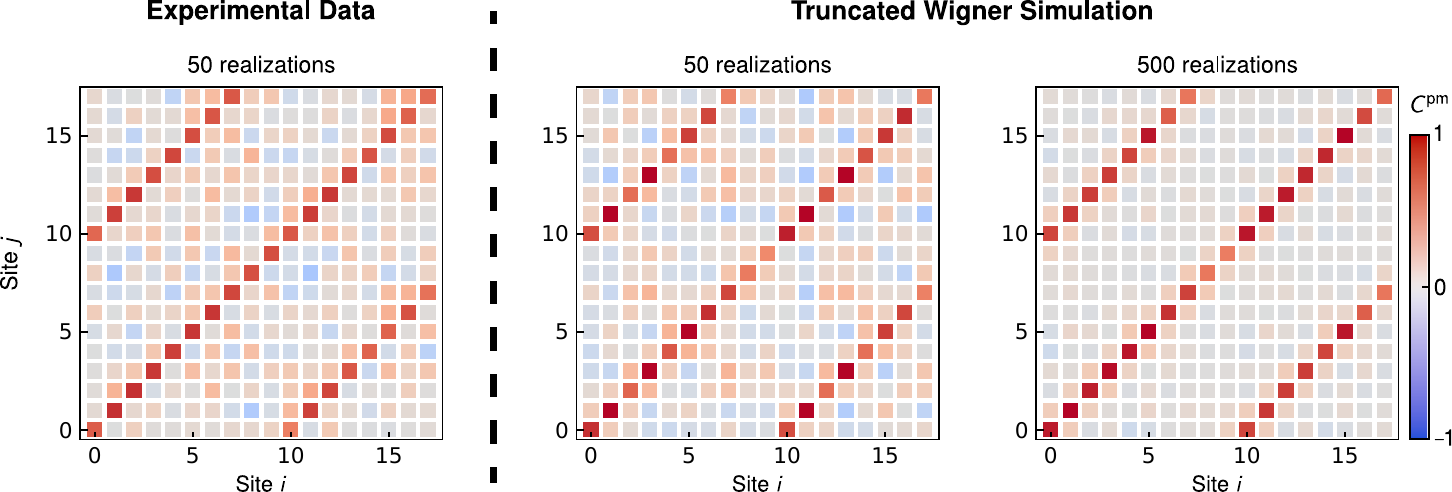}
    \caption{\textbf{Effect of finite statistics.} Left: correlation plot reproduced from Fig.~\ref{fig:overview}, showing $C^\mathrm{pm}$ obtained from 50 realizations of the experiment with interactions at distance $r = 10$. Right: simulation results obtained from a truncated Wigner approximation, where we either choose the same number of realisations as in the experiment or increase the number of realisations by a factor of 10 to reduce statistical uncertainty. The simulations indicate that residual correlations in the experimental data are mainly due to the finite sample size.}
    \label{ExtFig:FiniteStat}
\end{figure}
\newpage
\begin{figure}[h]
    \centering
    \includegraphics{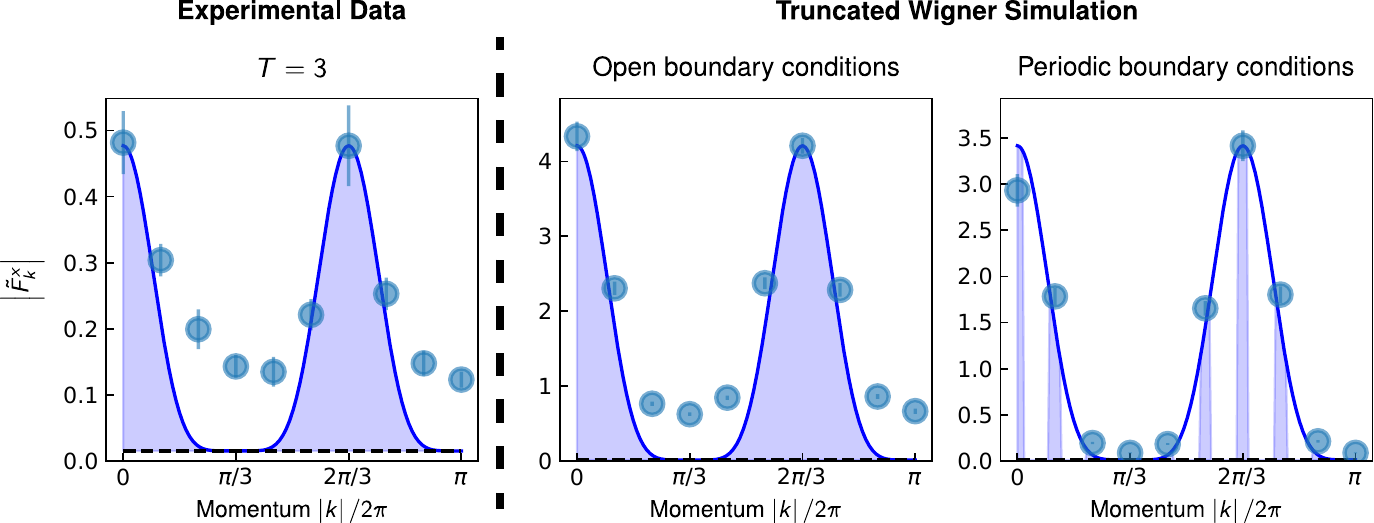}
    \caption{\textbf{Comparison between measured structure factor and simulation results.} The left graph shows the measured structure factor after $T=3$ Bloch periods of evolution, which is also shown in Fig.~\ref{fig:sideband_frequency}c. The two plots at right show results of a truncated Wigner simulation with and without periodic boundary conditions. In the first case, we find that the simulation has a similar offset as the experimental data with respect to the theoretical prediction (blue line). We attribute this offset to the finite system size, as the model is exact only for an infinite system or a system with periodic boundary conditions. Repeating the same simulation with a pulsed drive shown on the right shows that in this case the simulation data is consistent with the analytical model.}
    \label{ExtFig:Comp_Sim_Data}
\end{figure}
\clearpage
\newbibstartnumber{56}

\subfile{AuxTexFiles/supplement}

\end{document}

%% file: AuxTexFiles/preamble.tex
\usepackage{graphicx,subfigure}
\usepackage{amsmath,amssymb}
\usepackage{mathtools}
\usepackage{multirow}
\usepackage{textcomp}
\usepackage{float}
\usepackage{xcolor}
\usepackage{ulem}
\usepackage{dsfont}
\usepackage{tabularx}
\usepackage{comment}
\usepackage{subfiles}

\usepackage{bibunits}
\usepackage{natbib}
\defaultbibliographystyle{apsrev4-1}
\defaultbibliography{programmable}

\usepackage{etoolbox}

\makeatletter
\newcommand*{\newbibstartnumber}[1]{%
  \apptocmd{\thebibliography}{%
    \global\c@NAT@ctr #1\relax
    \addtocounter{NAT@ctr}{-1}%
  }{}{}%
}
\makeatother

\newcolumntype{Y}{>{\center\arraybackslash}X}

\usepackage{hyperref}
\hypersetup{colorlinks=true, urlcolor = blue}

\newcommand{\avg}[1]{\left\langle #1 \right\rangle}

\newcommand{\ket}[1]{\ensuremath{\left\vert{#1}\right\rangle}}

\newcommand{\vc}[1]{\ensuremath{\mathbf{#1}}}

\newcommand{\abs}[1]{\ensuremath{\left\vert{#1}\right\vert}}

\newcommand{\micro}[1]{\ensuremath{\mu\mathrm{#1}}}

\renewcommand{\micro}[1]{\ensuremath \mu\mathrm{#1}}

\renewcommand{\vec}[1]{\ensuremath{\mathbf{#1}}}

\newcommand{\Cov}[2]{\ensuremath{\mathrm{Cov}(#1,#2)}}
\newcommand{\Corr}[2]{\ensuremath{\mathrm{Corr}(#1,#2)}}
\newcommand{\Var}[1]{\ensuremath{\mathrm{Var}(#1)}}

\newcommand{\Cxx}{\ensuremath{C^{xx}}} 

\newcommand{\lev}{\ensuremath{a}}
\newcommand{\Ftil}[2][x]{\ensuremath{\tilde{F}^{#1}_{#2}}}
\newcommand{\mat}[1]{\mathbf{#1}}
\newcount\colveccount

\newcommand*\colvec[1]{
	\global\colveccount#1
	\begin{pmatrix}
		\colvecnext
	}
	\def\colvecnext#1{
		#1
		\global\advance\colveccount-1
		\ifnum\colveccount>0
		\\
		\expandafter\colvecnext
		\else
	\end{pmatrix}
	\fi
}

%% file: AuxTexFiles/end_matter.tex
We thank S. Gubser for illuminating discussions that inspired our exploration of non-Archimedean geometry.  We also acknowledge stimulating discussions with G. Bentsen, A. Daley, I. Bloch, B. Lev, N. Berloff, A. Deshpande, B. Swingle, and P. Hayden.  This work was supported by the DOE Office of Science, Office of High Energy Physics and Office of Basic Energy Sciences under Grant No. DE-SC0019174.  A. P. and E. S. C. acknowledge support from the NSF under Grant No. PHY-1753021. We additionally acknowledge support from the National Defense Science and Engineering Graduate Fellowship (A. P.), the NSF Graduate Research Fellowship Program (E. J. D. and E. S. C.), the Hertz Foundation (E. J. D.), and the German Academic Scholarship Foundation (J. F. W.).

\section*{Author Information}
Avikar Periwal, Eric S.~Cooper, and Philipp Kunkel contributed equally.

\section*{Author contributions}
A.~P., E.~S.~C., P.~K., J.~F.~W., and E.~J.~D. performed the experiments.  A.~P., E.~S.~C., P.~K., and M.~S.-S. analyzed the experimental data and developed supporting theoretical models.  A.~P., E.~S.~C., P.~K., and M.~S.-S. wrote the manuscript. All authors contributed to the discussion and interpretation of results.

%% file: AuxTexFiles/supplement.tex
\onecolumngrid
\begin{bibunit}
\begin{center}
 {\Large\textbf{Supplementary Information}}
\end{center}
\vspace*{.5cm}
\renewcommand{\figurename}{\textbf{Fig.}}
\renewcommand{\theequation}{S\arabic{equation}}
\renewcommand{\thefigure}{S\arabic{figure}}
\setcounter{equation}{0}
\setcounter{figure}{0}
\setcounter{section}{0}
This supplement provides further details about the derivation of the models and analysis methods used for our experiment.
In Sec. I, we derive the effective Hamiltonian in both position and momentum space representations. In Sec. II, we elaborate on the geometry reconstruction. In Sec. III, we describe our simulations based on the truncated Wigner approximations and provide additional comparisons between experimental data and simulation results.

\section{Hamiltonian Engineering}

In this work, we implement a family of translationally invariant XY spin models with effective interaction Hamiltonians of the form
\begin{equation}
H_I = -\sum_{\mu\nu} J(r_{\mu \nu}) f^+_\mu f^-_\nu,
\end{equation}
where the coupling $J(r_{\mu \nu})$ depends on the distance $r_{\mu \nu}$ between spins $\vec{f}_\mu$  and $\vec{f}_\mu$. For our system, $f=1$, and we work in units where $\hbar = 1$. The implementation of these models builds on our demonstration of long-range photon-mediated spin-exchange interactions in Ref. \cite{davis2019photon}.  Our approach, following the proposal of Ref. \cite{hung2016quantum}, is to apply a magnetic field gradient $2\omega_B/\mu_B$ per ensemble spacing,  introducing an energy cost $\omega_B r_{\mu\nu}$ for a flip-flop of spins $f_\mu$ and $f_\nu$, where $\mu_B$ is the Bohr magneton.  To turn on interactions at a distance $r$, we then require a pair of control fields differing in frequency by $\omega_B r$.  More generally, the frequency spectrum of the drive laser dictates the structure of the couplings $J(r)$.

\subsection{Derivation of the Effective Hamiltonian}

In the absence of a magnetic field gradient, the optical cavity in our system mediates interactions between all pairs of atoms, irrespective of the distance between them \cite{davis2019photon,davis2020protecting}. For a magnetic field oriented perpendicular to the cavity axis, interactions are well described by an all-to-all spin-exchange Hamiltonian. We approximate all spins as uniformly coupled to the cavity and parameterize the atom-light interaction by the vector light shift $\Omega$ produced by a circularly polarized intracavity photon. In this limit, the interaction Hamiltonian and spin exchange couplings are
\begin{align}
    H_I(t) &=\tilde{J}_+ (t) F^+ F^- + \tilde{J}_-(t) F^- F^+, \\
    \tilde{J}_\pm(t) &= \frac{\bar{n}_\text{ph}(t)\Omega^2}{4}\frac{\delta_\pm}{\delta_\pm^2+\kappa^2},
\end{align}
where $\boldsymbol{F} = \sum_\mu {\mathbf{f}_\mu}$ is the collective spin for all atoms, $\kappa$ is the cavity linewidth, and $\bar{n}_\text{ph}(t)$ is the instantaneous intracavity photon number.  The strength and sign of interaction depend on the detunings $\delta_\pm = \delta_c \mp \omega_z$ from two-photon resonances that flip a single spin, given in terms of the detuning $\delta_c$ of the drive field from cavity resonance and Zeeman splitting $\omega_z$~\cite{davis2019photon, davis2020protecting}. 

The interaction Hamiltonian can be rewritten in terms of a single interaction strength $\tilde{J}(t) = -(\tilde{J}_+ + \tilde{J}_-)$ as
\begin{equation}\label{eq:Jtilde}
\begin{aligned}
    H_I  &=-\tilde{J}(t) F^+ F^- - 2\tilde{J}_-(t) F^z,
\end{aligned}
\end{equation}
where we have defined $\tilde{J}$ to be positive for ferromagnetic interactions.  The final term in Eq.~\eqref{eq:Jtilde} acts as a uniform field along $\hat{z}$ and is suppressed by a factor of $n$ as compared to the collectively enhanced interactions between ensembles of $n$ atoms. This term can always be ignored by operating in a suitable rotating frame, and in our case its value of $2\tilde{J}_- = 2\pi \times 0.2\,\text{Hz}$ is negligibly small because it is not collectively enhanced.

To obtain a localized effective Hamiltonian that supports pair creation dynamics, we include the terms for a magnetic field gradient proportional to $\omega_B$ and quadratic Zeeman shift $q$. The full time-dependent Hamiltonian for our array of atomic ensembles is then
\begin{equation}
H(t) = -\sum_{l,m} \tilde{J}(t) F^+_l F^-_m  + \sum_l l \omega_B F^z_l + q\sum_\mu (f^z_\mu)^2,
\label{eq:HtFull}
\end{equation}
where $\vc{F}_l = \sum_{r_{\mu} = l} \vc{f}_\mu$ represents the collective spin on each site $l$. Viewing each site in a frame rotating at the local Larmor frequency $l\omega_B$, we can recast the Hamiltonian as
\begin{equation}
H(t) = -\sum_{l,m}  \tilde{J}(t) e^{i \omega_B (l-m)t}F^+_l F^-_m + q\sum_\mu \left(f^z_\mu\right)^2.
\label{Seq:HtRot}
\end{equation}

For sufficiently weak couplings that are modulated periodically at harmonics of the Bloch frequency $\omega_B$, with $n \tilde{J}(t), q < \omega_B$, the Floquet Hamiltonian for the system is well approximated by the time-averaged Hamiltonian, which is the lowest order term of the Floquet-Magnus expansion \cite{bukov2015universal},
\begin{equation}
H_\text{eff} = \frac{1}{\tau_B} \int_0^{\tau_B} dt\, H(t) = -\sum_{l, m} J(l-m) F^+_l F^-_{m} + q\sum_l (f^z_l)^2 = H_I + H_q.
\label{Seq:H_eff}
\end{equation}
Here we have defined $J(r)= J(l-m)$ using the Fourier transform of $\tilde{J}(t)$ evaluated at frequency $r\omega_B$,
\begin{equation}
J(r) = \frac{1}{\tau_B} \int_0^{\tau_B} dt\, e^{i r \omega_B t} \tilde{J}(t). 
\end{equation}
We can realize arbitrary couplings $J(r)$, subject to the hermiticity condition $J(r) = J^*(-r)$, by using the drive waveform

\begin{equation}
\tilde{J}(t) = \sum_{r} e^{-i r \omega_B t} J(r).
\end{equation}

\subsection{Momentum-Space Representation}

Since our scheme generates translationally invariant interactions, it is natural to write the Hamiltonian in momentum space, in terms of spin-wave operators
\begin{equation}
\tilde{\vc{F}}_k = \frac{1}{\sqrt{M}}\sum_{j=1}^M e^{-i k j} \vc{F}_j.
\end{equation}
The translation invariance is exact in the limit of a large system ($M\rightarrow \infty$) or for a drive chosen to induce periodic boundary conditions, $J(r) = J(M-r)$.  For these cases, we can rewrite Eq.~\eqref{Seq:H_eff} in terms of spin-wave operators to obtain the momentum-space representation of the Hamiltonian
\begin{equation}
H = -\sum_{k}\tilde{J}_k \Ftil[+]{-k} \Ftil[-]{k} + q\sum_\mu f_\mu^2,
\end{equation}
with momentum-space couplings given by
\begin{equation}
\tilde{J}_k = \sum_{r=0}^{M-1} e^{-i k r} J(r) = \frac{1}{\tau_B}\int_0^{\tau_B}dt\, \delta_M(\omega_B t - k) J(t) \rightarrow \tilde{J}(t = k/\omega_B),
\end{equation}
where $\delta_M(x) \equiv (1-e^{iMx})/({1-e^{ix}})$ tends to a Dirac delta function in the limit of an infinite system.

We initialize the system with all atoms in $m=0$, and for early times we can approximate the atoms in this state as a constant classical pump field that drives the formation of correlated atom pairs in states $m=\pm 1$ \cite{davis2019photon}.  In this regime, the Hamiltonian can be expressed in terms of bosonic operators $a^\dagger_i$ ($b^\dagger_i$) representing the creation of an atom in state $m=1$ ($m=-1$) on site $i$ (see Eq.~\ref{eq:Fbosonic} below) or, alternatively, the momentum-space counterparts $a^\dagger_k, b^\dagger_k$.  In terms of these bosonic operators, the spin-wave operators can be written as
\begin{equation}
\begin{aligned}
\tilde{F}_{-k}^+ &= \sqrt{2n} \left(a_k^\dagger + b_{-k} \right), \\
\tilde{F}_k^- &= \sqrt{2n} \left(a_k + b_{-k}^\dagger \right),
\end{aligned}
\end{equation}
and the interaction Hamiltonian is
\begin{equation}
    H_I = \chi_k \sum_k [a_k^\dagger a_k + a_k^\dagger b_{-k}^\dagger + b_{-k}a_k + b_{-k}b_{-k}^\dagger].
\end{equation}
This shows that $\chi_k = -2n\tilde{J}_k$ is the dispersion relation for spin excitations. This is identical to the dispersion relation obtained from a Holstein-Primakoff transformation in the vicinity of a spin-polarized state. 

\subsection{Early-Time Dynamics}
\label{sec:EarlyDynamics}
In this section we present an analytical derivation of the dynamics of our system at early times, when we can neglect depletion of atoms in $m=0$. In contrast to the previous sections, the derivations here do not require the limit of weak interactions and capture the discretized nature of the dynamics over each Bloch period. Our approach will be to express the full time-dependent Hamiltonian in terms of spin-wave operators. Solving for the dynamics of the spin waves facilitates the computation of the structure factor and spatial correlations.
We start with the Hamiltonian in the rotating frame as written in Eq.~\eqref{Seq:HtRot}. We recognize that the interaction term can be written succinctly in terms of spin-wave operators $\tilde{\mathbf{F}}_k$. The full expression for this Hamiltonian is
\begin{equation}
H(t) = -M \tilde{J}(t) \Ftil[+]{k=-\omega_B t} \Ftil[-]{k=\omega_B t} + H_q,
\end{equation}
where $H_q = q \sum_\mu (f^z_\mu)^2$ is the quadratic Zeeman shift. 
Since the system is finite, there are only $M$ orthogonal momentum modes, whereas the cavity couples to a continuous range of momenta $k=\omega_B t$ during a single Bloch period. Generically this couples adjacent momentum modes. In the limit of an infinite system, the modes become fully independent.

The modes are also fully independent in the case where we generate periodic boundary conditions using a pulsed drive field.  In this case, the instantaneous spin-exchange coupling $\tilde{J}(t) = \sum_k \frac{2\pi}{M} \tilde{J}_k \delta(\omega_Bt-k)$ only takes on non-zero values $M$ times per Bloch period, and we observe that the momentum modes decouple in the Hamiltonian
\begin{equation}
H(t) = -2\pi \sum_k \tilde{J}_k \delta(\omega_Bt-k) \Ftil[+]{_k} \Ftil[-]{k} + H_q.
\label{Seq:PulsedH}
\end{equation}
For any given momentum mode, the evolution is discrete, with a short period of coupling to the optical cavity followed by a longer period of time when the state is acted upon by the quadratic Zeeman shift. In the limit of weak interactions and many Bloch periods, this trotterized Hamiltonian becomes identical to $M$ independent continuous pair-creation processes in different momentum modes. 

In order to analytically solve for the dynamics of pair creation, we define three bosonic modes on each site $a_i$, $b_i$, $c_i$ corresponding to atoms on site $i$ in the states $m = \pm 1,0$ respectively. For early times we can treat the atom initialized in $m=0$ as a classical pump field with $c_i = \sqrt{n}$. In this limit we can  express the local spin operators as a function of operators for the bosonic modes,
\begin{equation}\label{eq:Fbosonic}
F_i^+ = (F_i^-)^\dagger = \sqrt{2}  \left(a_i^\dagger c_i + c_i^\dagger b_i \right) =  \sqrt{2n} \left(a_i^\dagger + b_i \right),
\end{equation}
and similarly express the quadratic Zeeman shift as
\begin{equation}
H_q = q \sum_i \left(a_i^\dagger a_i + b_i^\dagger b_i\right).
\end{equation}
Following the same convention as for the spin-wave operator $\tilde{F}_k$, we define operators for the momentum modes $a_k \equiv \frac{1}{\sqrt{M}} \sum_l e^{-ikl} a_l$ and $b_k \equiv \frac{1}{\sqrt{M}} \sum_l e^{-ikl} b_l$. We can rewrite the key operators in the rotating-frame Hamiltonian (Eq. \eqref{Seq:PulsedH}) in terms of these bosonic modes,
\begin{equation}
\begin{aligned}
\tilde{F}_{-k}^+ &= \sqrt{2n} \left(a_k^\dagger + b_{-k} \right) \\
\tilde{F}_k^- &= \sqrt{2n} \left(a_k + b_{-k}^\dagger \right) \\
H_q &= q \sum_k \left(a_k^\dagger a_k + b_k^\dagger b_k\right).
\end{aligned}
\end{equation}

With the Hamiltonian expressed in terms of the bosonic operators, we can proceed to solve for the dynamics of the system. In particular, we realize that there are independent equations of motion for every discrete value of $k = 2\pi m/M$ where $m$ is an integer. In particular, the Heisenberg equations of motion for $a_k$ and $b_{-k}^\dagger$ depend only on each other,
\begin{subequations}
\label{Seq:MomentumSpaceEOM_UnAveraged}
\begin{align}
    \dot{{a}}_k &= i[H,{a}_k] = -2\pi i \chi_k \delta(\omega_Bt - k) \left({a}_{k} + {b}_{-k}^\dagger\right) - i q  a_k
    \\
    \dot{{b}}_{-k}^\dagger &= i[H,{b}_{-k}^\dagger] = 2\pi i \chi_k \delta(\omega_Bt - k) \left({a}_{k} + {b}_{-k}^\dagger\right) + i q  b^\dagger_{-k},
\end{align} 
\end{subequations}
where $\chi_k = -2n\tilde{J}_k$ is the dispersion relation. To solve these equations, we write the two operators as a vector $\mathcal{O} = (a_{k}, b_{-k}^\dagger)$. This leads to a matrix equation $\dot{\mathcal{O}} = D(t) \mathcal{O}$ where,
\begin{equation}
D(t) = 2\pi i\chi_k \delta(\omega_Bt - k) 
 \begin{bmatrix}
    -1 & -1  \\
   1 & 1
\end{bmatrix}
+ iq 
 \begin{bmatrix}
    -1 & 0 \\
    0 & 1
\end{bmatrix}.
\end{equation}
Integrating the equations of motion over a single Bloch period starting from the first interaction pulse at $t = k/\omega_B$ yields a propagation matrix $\Pi_k = QX_k$ where the propagation due the quadratic Zeeman shift is 
\begin{equation}
    Q =
 \begin{bmatrix}
    e^{-iq\tau_B} & 0 \\
    0 & e^{iq\tau_B}
\end{bmatrix}
\end{equation}
and the propagation for the short interaction pulse is
\begin{equation}
\begin{aligned}
X_k &=  \begin{bmatrix}
    1 & 0  \\
   0 & 1
\end{bmatrix} + i\chi_k\tau_B
\begin{bmatrix}
    -1 & -1  \\
   1 & 1
\end{bmatrix}, 
\end{aligned}
\end{equation}
where $\tau_B = 2\pi/\omega_B$ is the Bloch period for spin waves. The propagator $\Pi_k$ can be diagonalized using Bogoliubov modes, 
\begin{equation}
u_\pm = \frac{-e^{-iq\tau_B}}{\chi_k\tau_B}\left[\sin(q\tau_B) + \cos(q\tau_B)\chi_k\tau_B \pm i \sqrt{-1+\left(\cos(q\tau_B)-\chi_k\tau_B\sin(q\tau_B)\right)^2 } \right] a_k + b_{-k}^\dagger,
\end{equation}
and the corresponding eigenvalues are
\begin{equation}
\lambda_\pm = \cos(q\tau_B) - \chi_k\tau_B \sin(q\tau_B) \pm \sqrt{-1+\left(\cos(q\tau_B)-\chi_k\tau_B\sin(q\tau_B) \right)^2 }.
\end{equation}

For our system it is useful to take the limit of large interaction strength to simplify the above expressions. In the limit of strong instability where $\chi_k <0 $ and $q >0$ have opposite signs and $|\chi_k|\tau_B\sin(q\tau_B) > 1$, the eigenvalues of the propagator tend to $\lambda_- = 0$ and $\lambda_+ = 2\cos(q\tau_B)+ 2|\chi_k|\tau_B\sin(q\tau_B)$. From this we determine that amplitude of momentum mode $k$ grows at a rate set by the dispersion relation $\chi_k$, with the minimum of the dispersion $\chi_k < 0$ corresponding to the maximum growth rate.

\subsection{Structure Factor}
We directly observe the growth of momentum modes through measurements of the structure factor. The squared magnitude of the structure factor $|\Ftil[x]{k}|^2$ can be directly expressed in terms the bosonic operators $a_{\pm k}, b_{\pm k}$ that we computed in section \ref{sec:EarlyDynamics}. When the $m=0$ pump mode can still be treated classically, the magnitude of the structure factor is
\begin{equation}
\begin{aligned}
|\Ftil{k}|^2 &= 
\avg{ \Ftil{k} \Ftil{-k}} =
\frac{1}{4}\avg{\left(\Ftil[+]{k} + \Ftil[-]{k}\right) \left(\Ftil[+]{-k}+\Ftil[-]{-k} \right)} \\
&=\frac{1}{4}\avg{\Ftil[+]{k}\Ftil[-]{-k}} + \frac{1}{4}\avg{\Ftil[-]{k}\Ftil[+]{-k}} \\
&= \frac{n}{2} \left\langle \left(a_{-k}^\dagger + b_{k} \right)\left(a_{-k} + b_{k}^\dagger \right) \right\rangle + \frac{n}{2}\left\langle \left(a_{k} + b_{-k}^\dagger \right)\left(a_{k}^\dagger + b_{-k}\right) \right\rangle.
\label{Seq:structure_factor_modes}
\end{aligned}
\end{equation}

 We determine the final values of the operators $a_k,b^\dagger_{-k}$ in terms of their initial values by using the relation $\mathcal{O}(T\tau_B) = \Pi^{T}\mathcal{O}(0)$. 
 Evaluating the norm using the vacuum state after a single Bloch period, each of the two terms for the squared magnitude of the structure factor may be computed exactly as
 \begin{equation}
    \frac{1}{4}\avg{\Ftil[\mp]{ k}\Ftil[\pm]{-k}} = \frac{n}{2} \left[1-4\chi_{\pm k}\tau_B \cos(q\tau_B)\sin(q\tau_B) + \left(2\chi_{\pm k}\tau_B\right)^2\sin(q\tau_B)^2\right].
\end{equation}
For sufficiently large interaction strength, we can obtain a simple expression for $\avg{\Ftil[\mp]{ k}\Ftil[\pm]{-k}}$ after multiple Bloch periods. To leading order in $|\chi_k|$, growth over a single Bloch period is given by $|\lambda_{+}(\pm k)|^2$ where $\lambda_{+} (\pm k) \approx 2|\chi_{\pm k}|\tau_B\sin(q\tau_B)$. The structure factor is thus given approximately by
\begin{equation}
|\Ftil{k}|^2 = \frac{n(2\tau_B)^{2T}}{2} \left(\chi_k^{2T}+\chi_{-k}^{2T}\right)\sin(q\tau_B)^{2T}.
\end{equation}
If we consider only drive waveforms $\tilde{J}(t)$ with phases of $0$ or $\pi$ -- as is true for all waveforms used in this paper -- we have $\tilde{J}(t) = \tilde{J}(-t)$. This simplifies the expression for the magnitude of the structure factor to
\begin{equation}
    |\Ftil{k}| = \sqrt{n} |\chi_{k}|^{T}\sin(q\tau_B)^{T}\propto \tilde{J}(k/\omega_B)^T.
\end{equation}

\section{Geometry Reconstruction}
\subsection{Euclidean Reconstruction}
In this section we give an overview of classical metric multidimensional scaling~\cite{torgerson1952multidimensional}, the method used for extracting coordinates in the effective geometries presented in Figs.~\ref{fig:geometry}-\ref{fig:trees} of the main text. To calculate distances between sites, we begin at the Gaussian ansatz $\Cxx_{ij} = \exp(-ad_{ij}^2),$ where $d_{ij}$ is the inferred distance between sites $i$ and $j$.  
The free parameter $a$ is chosen so that the strongest correlations correspond to a distance $d_{ij}=1$.  The normalization of $\Cxx$ enforces a distance of 0 from a site to itself.  
We assume a linear translational invariance of the system, not including periodic boundary conditions, so that the $M(M - 1)/2$ correlation measurements are used to numerically fit $M - 1$ possible distances and one free parameter $a$.  
From the $M - 1$ distances we construct an effective distance matrix, where the distance is constant along each diagonal.  We let $\mat{D}$ be the $M\times M$ matrix of pairwise squared distances. From $\mat{D}$, we wish to calculate an $M\times k$ matrix $\boldsymbol{\rho}$, corresponding to $k$-dimensional coordinates for each of the $M$ sites.

It is instructive to work backwards from a set of coordinates to calculate the expected squared distance, denoted by $\mat{D}$.  To that end, we have
\begin{equation}
\label{eq:distance_from_coordinates}
\begin{aligned}
  \mat{D}_{ij} &= \sum_k (\boldsymbol{\rho}_{ik} - \boldsymbol{\rho}_{jk})^2\\
          &= \sum_k\left( \boldsymbol{\rho}_{ik}\boldsymbol{\rho}^{\intercal}_{ki} + \boldsymbol{\rho}^{\intercal}_{kj}\boldsymbol{\rho}_{jk} - 2\boldsymbol{\rho}_{ik}\boldsymbol{\rho}^{\intercal}_{kj}\right).
\end{aligned}
\end{equation}
We recognize the last term as the matrix product $\boldsymbol{\rho}\boldsymbol{\rho}^{\intercal}$, and the first two terms as constants added to each row and column of $\mat{D}$.  In order to isolate the $\boldsymbol{\rho}\boldsymbol{\rho}^{\intercal}$ term we define the centering matrix $\mat{C} = \mat{I} - 1/M$.  $\mat{C}$ acts on an $M$-dimensional vector by subtracting out its mean.  We choose the coordinates to be centered around the origin, so that when we left- and right-multiply Eq.~\eqref{eq:distance_from_coordinates} by $\mat{C}$, the first two terms vanish.  The last term remains unchanged, and we are left with
\begin{equation}
-\frac{1}{2}\mat{C}\mat{D}\mat{C}
  = \boldsymbol{\rho}\boldsymbol{\rho}^{\intercal}.
\end{equation}

If $\boldsymbol{\rho}$ is a set of coordinates in $k$ dimensions, then $\mat{C}\mat{D}\mat{C}$ must have rank $k$.  To enforce this, we take the truncated singular value decomposition, writing $-\mat{C}\mat{D}\mat{C} = 2\mat{U}\mat{\Sigma}\mat{V},$ with $\mat{U}, \mat{V}$ unitary and $\mat{\Sigma}$ diagonal with $k$ elements.  As a distance matrix must be square, $\mat{U} = \mat{V}$, and we have $\boldsymbol{\rho} = \mat{U}\mat{\Sigma}^{1/2}$.  This metric approach is equivalent to a principal component analysis of the centered distance matrix, and metric multidimensional scaling is also referred to as principal coordinates analysis~\cite{gower2014principal}.

\subsection{Bulk Reconstruction Motivation}
In the Euclidean geometry reconstruction, we inferred a metric of the system by calculating the effective distances between sites and the paths connecting them.  In reconstructing the bulk we take a different approach, and aim to find a symmetry group corresponding to the bulk via measurements on sites that we assume are on the boundary of the system. We determine the locations of the sites on the boundary theory via the multidimenensional scaling algorithm described in the prior section, based on correlations $\Cxx$.  Since the coordinates $\boldsymbol{\rho}$ must be centered, and the Hamiltonians we work with have complete translational invariance, the calculated coordinates inevitably lie on a natural boundary. Within this reconstruction, strongly correlated sites will be close together on the boundary.

Mathematically, any geometry has an associated symmetry group.  For example, in a square the associated group is $S_4$, whose elements are all the permutations of the 4 vertices.  In Minkowski space, the associated symmetry group is the Lorentz group, which is generated by the boosts and rotations of special relativity.  For the $p$-adic AdS/CFT correspondence, the treelike gravitational bulk is specified by the Bruhat-Tits tree~\cite{heydeman2016tensor, gubser2017p}.

Cayley's theorem states that any group can be written in terms of permutations between sites, and in our discrete system these are natural operations.  The choice of permutations is motivated by the Ryu-Takayanagi formula. This conjecture states that for any region $A$ on the boundary CFT, the entanglement entropy is proportional to the minimal area surface with boundary $A$ in the AdS bulk~\cite{ryu2006holographic}.  Sites that are highly correlated with each other, and not strongly correlated with the rest of the system, must correspond to a small bulk area.  

We begin the bulk reconstruction by finding the physical distance $r=\abs{i-j}$ such that $C^{xx}(r)$ is maximized, drawing a bond between each pair of sites $(i, i\pm r\bmod N)$ separated by this distance.  The path between these two sites should correspond to a minimal area within the bulk.  We draw connections in two directions, because we assume that the sites live on a closed boundary, and interactions between the sites must obey periodic boundary conditions.  Each bond $(i, j)$ corresponds to the permutation of sites $i$ and $j$, and the set of bonds generates a subgroup of the geometry's overall symmetry group.

The crux of the bulk reconstruction is the iterative step, where we treat each bond as a new site, and repeat the process with a coarse-grained version of the initial correlation matrix.  This can be thought of as taking the quotient of the unknown symmetry group with the subgroup generated by the previously drawn bonds, so that we can uncover another subgroup.  In addition to the group theoretic motivation, the coarse-graining is physically motivated.  In the AdS/CFT correspondence, the emergent dimension of the gravitational bulk captures the renormalization group of the boundary field theory, with high energy on the boundary and low energy on the bulk interior~\cite{hartnoll2018holographic,qi2018does}.  The coarse-graining step acts as a renormalization step, moving from high energy  (corresponding to short length scales in the reconstructed geometry) to low energy (corresponding to longer length scales).  This process must be repeated until the full symmetry group is recovered and each point on the boundary is connected.

\section{Truncated Wigner Approximation}
While the analytical model already provides a good intuition for the expected dynamics in the experiments, we gain further insight by additionally comparing our results to a numerical simulation based on the truncated Wigner approximation (TWA).  This numerical simulation includes depletion effects of the state $m=0$ and additional experimental imperfections.  Instead of solving the full quantum evolution, the TWA relies on solving the classical mean-field equations of motion, which makes the simulation computationally tractable. Additionally, the statistical sampling of the TWA incorporates effects of finite statistics for a better comparison with the experimental results.

\subsection{General concept}

The truncated Wigner approximation is a semiclassical method designed to numerically solve the dynamics of a quantum system in a way that is computationally feasible. To this end, one simulates the mean-field equations of motion for different initial states in a classical phase space. These initial states are sampled from the Wigner distribution of the initial quantum state. Intuitively, the truncated Wigner approximation assumes that the quantum dynamics can be simulated by taking the quantum fluctuations of a given state as a statistical ensemble and propagating this ensemble according to the classical mean-field equations of motion. The statistical ensemble after propagation is then interpreted as the Wigner distribution of the time-evolved quantum state, from which one can extract the observables of interest.

This semiclassical approach, by construction, has some limitations, which are discussed in detail in Refs.~\cite{Sinatra2002,Blakie2008,Lewis-Swan2016}. Since it is difficult to derive a general limit for the validity of the truncated Wigner approach, we provide in the following some intuitive arguments for the validity regions:

First, since the Wigner function of the initial quantum state is treated as a statistical ensemble, the truncated Wigner approximation works best in situations where the Wigner distribution is positive. This excludes highly entangled quantum states as initial states. However, one is often interested in the time evolution of initial coherent states, which have a positive Wigner distribution and are, therefore, well suited for this approach.

Second, the pair-creation dynamics that we simulate generate a highly entangled two-mode squeezed state. For short evolution times, this leads to squeezing of the Wigner distribution, which is well captured by the truncated Wigner approximation. For longer evolution times, however, this approach fails to correctly represents the non-Gaussian entangled states that evolve. Yet, if one is mainly interested in the first and second moments of the Wigner distribution instead of the exact quantum state, the truncated Wigner approach still provides some insight into the evolution at these late times. Experimentally, we expect the late-time dynamics to be modified by dissipation due to photon loss and spontaneous emission, which can anyway remove the negativities of the Wigner function that the TWA fails to capture. Thus, the truncated Wigner approximation provides a useful tool for simulating the most relevant quantum dynamics, including saturation effects and technical imperfections.
\subsection{Mean-field Equations of Motion}
To derive the mean-field equations of motion, we start with the Hamiltonian
\begin{equation}
    \mathcal H = \sum_{i=1}^M\left[ -q\, n_{0,i} +i\omega_B\,{F}_{z,i}\right] + \tilde{J}(t)\sum_{i,j = 1}^M {F}^+_i {F}^-_j.
\end{equation}
Here, $M$ corresponds to the number of traps, $n_{0,i}$ is the atom number operator for atoms in the magnetic substate $m = 0$, $\omega_B$ is the magnetic field gradient, and $\tilde{J}(t)$ is the time-dependent coupling strength due to the modulated drive field.  To obtain the equation of motion for the annihilation (creation) operator of each magnetic substate at site $i$, we compute the commutator
\begin{equation}
    \frac{d{a}^{(\dagger)}_{i}}{dt} = i[\mathcal H,{a}^{(\dagger)}_{i}],
\end{equation}
 with analogous computations for ($b^{(\dagger)},c^{(\dagger)}$). To arrive at the the mean field equations, we make the substitution
\begin{equation}
    \hat{a}^{(\dagger)}_{i} \rightarrow \sqrt{n_i}\,\zeta^{(*)}_{1,i}, \qquad
    \hat{b}^{(\dagger)}_{i} \rightarrow \sqrt{n_i}\,\zeta^{(*)}_{-1,i}, \qquad
    \hat{c}^{(\dagger)}_{i} \rightarrow \sqrt{n_i}\,\zeta^{(*)}_{0,i},
\end{equation}
where $\zeta^{(*)}_{m,i}$ is a complex number whose absolute value squared $|\zeta^{(*)}_{m, i}|^2 = n_{m,i}/n_i$ corresponds to the fraction of atoms in the magnetic substate $m$ at site $i$.
Using matrix notation with 
\begin{equation}
    \vec{\zeta} = \colvec{5}{\zeta_{1,1}}{\zeta_{0,1}}{\zeta_{-1,1}}{\zeta_{1,2}}{\vdots}
\end{equation}
we can write the equations of motion for a linear gradient across the cloud as
\begin{equation}
    \frac{d\vec{\zeta}}{dt} = -i\,\left[\omega_B \mathcal{Z}-q\mathcal{Q}+2\tilde{J}(t)\mathcal{F}\right]\cdot\vec{\zeta},
\end{equation}
where the first two matrices are given by
\begin{equation}
  \mathcal{Z} =
  \begin{bmatrix}
    0 \cdot S_z & & \\
    &1\cdot S_z & &\\
    && \ddots & \\
    & & & n\cdot S_z
  \end{bmatrix},\quad
  \mathcal{Q} =
  \begin{bmatrix}
    (\mathbb{I}_3 -S_z^2) &  \\
    & \ddots & \\
    & & (\mathbb{I}_3 -S_z^2)
  \end{bmatrix},
\end{equation}
with
\begin{equation}
    S_z = \begin{bmatrix}
      1&0&0\\
      0&0&0\\
      0&0&-1
    \end{bmatrix}.
\end{equation}
The coupling matrix $\mathcal{F}$ is calculated via
\begin{equation}
    \mathcal F = \vec{\zeta}^+\cdot(\vec{\zeta}^+)^* + \vec{\zeta}^-\cdot(\vec{\zeta}^-)^*
\end{equation}
where $\vec{\zeta}^+$ and $\vec{\zeta}^-$ are given by 
\begin{equation}
    \vec{\zeta}^+ = \begin{bmatrix}
    0 & 1 &&&&&&\\
    & 0 & 1 &&&&&&\\
    && 0 & 0 &&&&&\\
    &&& \cdot &&&&&\\
    &&&& \cdot &&&&\\
    &&&&& \cdot &&&\\
    &&&&&& 0 & 1 &\\
    &&&&&&& 0 & 1\\
    &&&&&&&& 0
  \end{bmatrix}\cdot\vec{\zeta}
  ,\qquad
  \vec{\zeta}^- = \begin{bmatrix}
    0 &&&&&&&\\
    1 & 0 &&&&&&&\\
    & 1 & 0 &&&&&&\\
    &&& \cdot &&&&&\\
    &&&& \cdot &&&&\\
    &&&&& \cdot &&&\\
    &&&&& 0 & 0 &&\\
    &&&&&& 1 & 0 &\\
    &&&&&&& 1 & 0
    \end{bmatrix}\cdot\vec{\zeta}.
\end{equation}

The last step for implementing the truncated Wigner approximation is to sample the initial Wigner distribution. To simulate our experiments, we take as the initial state the one where all atoms are initialized in the magnetic substate $m = 0$, while the other two substates are empty. The quantum fluctuations of this state can then be modeled by adding a complex random number $X_{m,i}$ with variance $\Var{X_{m,i}} = 1/2$, i.e.
\begin{equation}
    \vec{\zeta}_0 =   \colvec{7}{0}{\sqrt{n_1}}{0}{}{0}{\sqrt{n_2}}{\vdots} + \colvec{7}{X_{1,1}}{X_{0,1}}{X_{-1,1}}{}{X_{1,2}}{X_{0,2}}{\vdots}.
\end{equation}
In our case, we choose the atom number on each site to be the same on average. By including the random numbers $X_{0,i}$ on the occupied modes, we are allowing for Gaussian fluctuations of the atom number on each site, resembling the experimental situation. One can verify that this sampling indeed reproduces the desired quantum fluctuations of the initial coherent state. For example, evaluating $\hat{F}^x_{i}$ on each site, we find the variance over multiple samples to be
\begin{equation}
    \Var{F^x_{i}} = \langle n_i\rangle,
\end{equation}
as expected for the initially prepared coherent state.
\subsection{Comparison with Experiment}
One of the main advantages of the truncated Wigner approach is that it allows for modeling the experimental situation by including saturation and technical fluctuations, such as shot-to-shot fluctuations of the magnetic field or of the power of the driving field. This helps to compare our measurements with theoretical predictions, including effects not captured by the analytical model.

\subsubsection{Geometry Reconstruction}
Here, we first use the TWA simulation to investigate the effects of finite statistics on the reconstruction of the geometry. As an example we choose the anti-ferromagnetic ladder, which is shown in Fig.~\ref{fig:geometry}\,c. In the experiment, we found that the reconstructed geometry was not simply a flat triangular lattice as one would naively expect but also exhibited some twisting in 3D. 

\begin{figure}
    \centering
    \includegraphics[width=\textwidth]{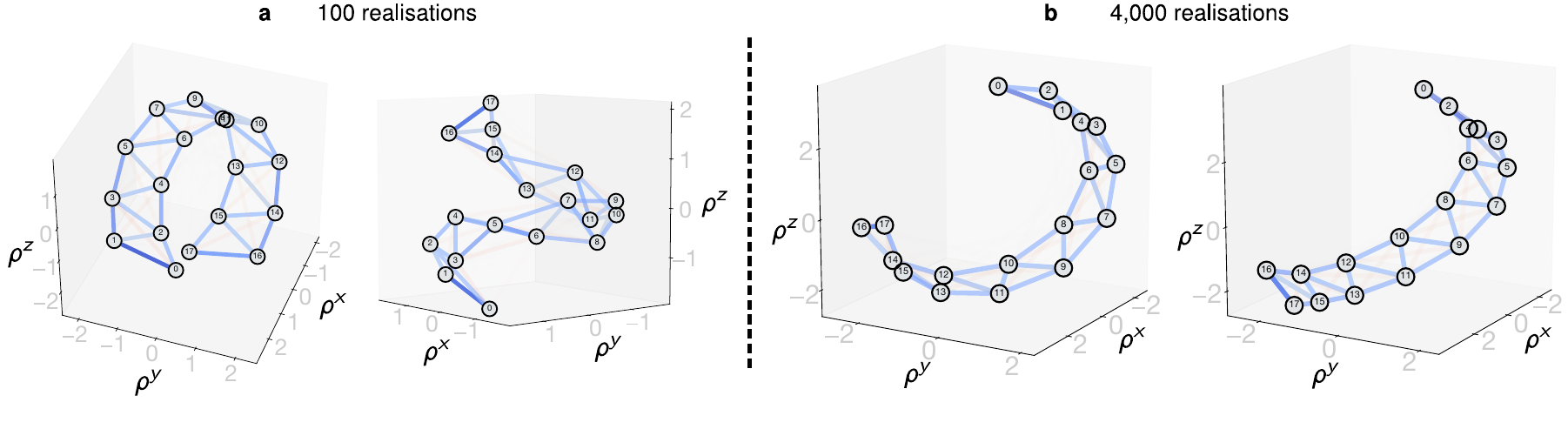}
    \caption{\textbf{Reconstructed Geometry for the anti-ferromagnetic ladder:} \textbf{a} shows two examples of the simulation results with 100 realizations. \textbf{b} shows two examples with 4,000 realizations each.}
    \label{Fig:SuppAFM_Ladder}
\end{figure}

For comparison, Fig.~\ref{Fig:SuppAFM_Ladder}a shows two examples of reconstructed geometries based on samples of 100 simulated realizations, a typical sample size for the reconstruction with the experimental data. Here, the finite statistics lead to a deformation of the structure due to the inherent fluctuations of the pair-creation process.  Compared with other graphs realized in our experiments, such as the cylinder and M\"{o}bius strip, the ladder is especially sensitive to these fluctuations because it is not constrained in 3D by the interactions.
Figure~\ref{Fig:SuppAFM_Ladder}b shows two examples for the corresponding TWA simulation with 4,000 realizations. In this case, the large fluctuations are reduced, but we still find a curved geometry instead of a planar lattice structure.  We attribute this bending to the finite statistics of the simulation, where the statistical fluctuations produce a small offset in the absolute value of the correlations even at a large distance. This small deviation from the Gaussian decay leads to a curved structure, since the effective distance between very weakly coupled sites is underestimated. We expect that the curvature decreases slower than logarithmically as a function of the number of realizations.

\subsubsection{Treelike Correlations}
\begin{figure}
    \centering
    \includegraphics[width=\textwidth]{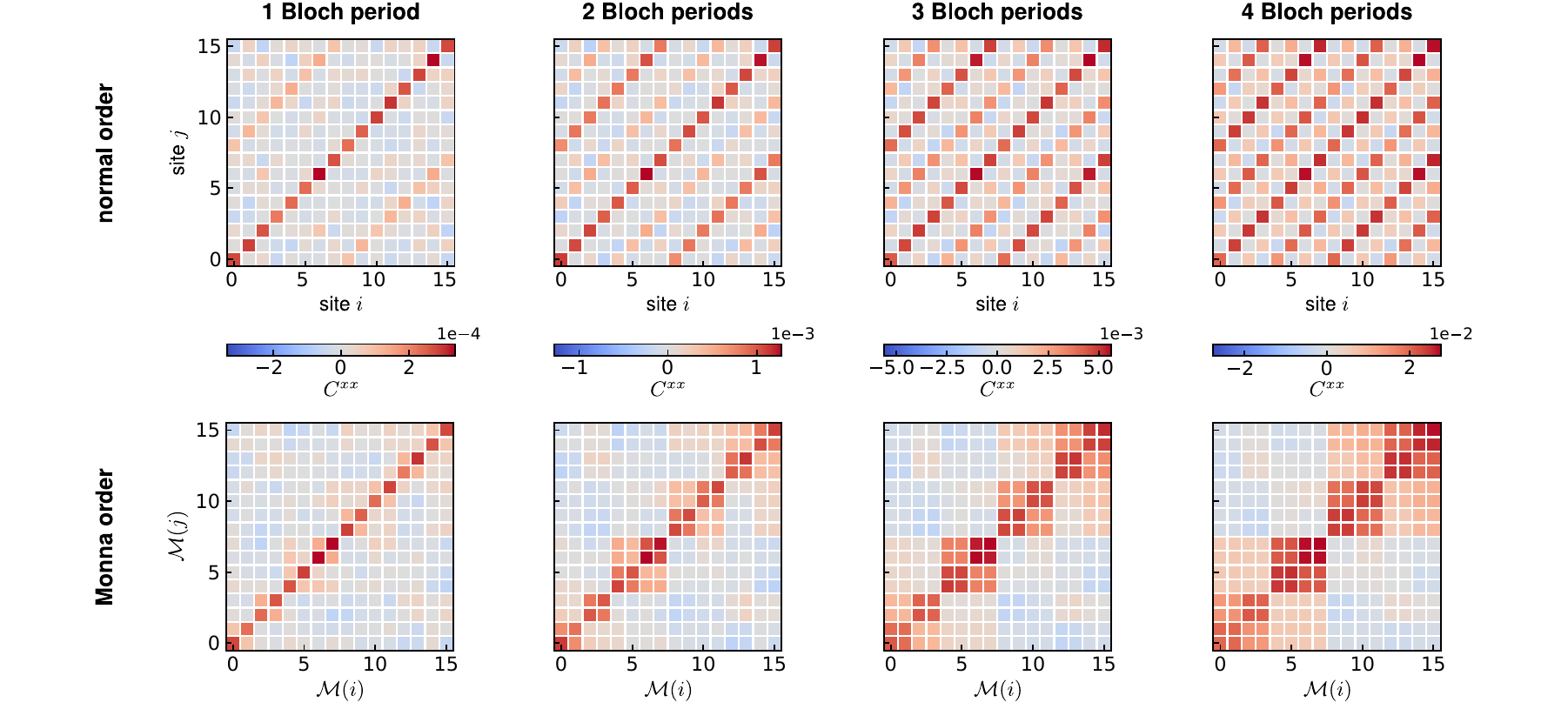}
    \caption{\textbf{TWA simulation of treelike correlation spreading.} The upper row depicts the spreading of correlations for the case of $s =1 $ as a function of time. The lower half shows the same data after rearranging the sites according to the Monna map. In the Monna-mapped case, one sees that the square structure builds up over time, which is a hallmark of the treelike geometry.}
    \label{Fig:Supp_Tree}
\end{figure}

Using the TWA simulation, we also study the time evolution of the system with non-Archimedean geometry, i.e. the model of Eq.~\eqref{eq:tree_couplings} with $s = 1$.  We compare the simulation to the experimental data shown in Fig.~\ref{fig:trees}b for a fixed evolution time of two Bloch periods. The time evolution of the simulated data is depicted in Fig.~\ref{Fig:Supp_Tree}. After rearranging the sites according to the Monna map, we find that the simulated spin-spin correlations begin to exhibit a block structure after 2 Bloch periods of evolution time, much as in the experiment. The simulation additionally shows that the complete block structure expected for a treelike geometry only becomes fully visible for longer evolution times. In particular, after two Bloch periods the $8\times 8$ block corresponding to correlations between sites with a physical distance $\abs{i-j}=1$ is not yet clearly visible, which is consistent with our experimental findings.

\subsubsection{Bipartite Correlations}
\begin{figure}
    \centering
    \includegraphics[width=\textwidth]{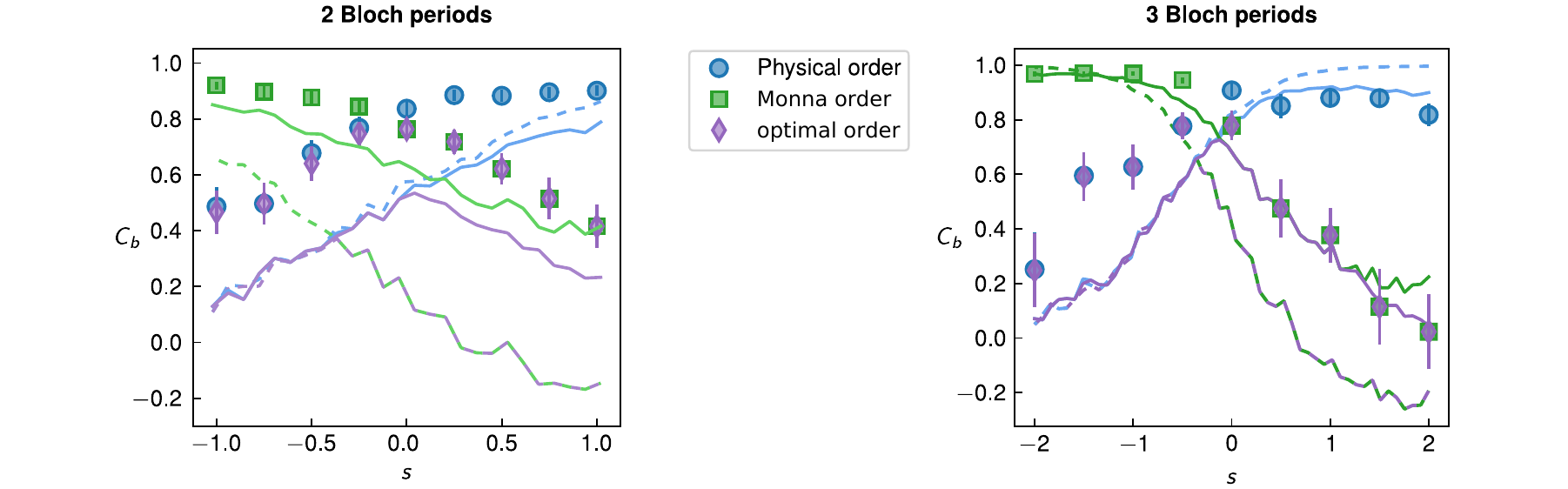}
    \caption{\textbf{Bipartite Correlations:} Comparison between the experimental data and simulation results for the bipartite correlations as functions of the parameter $s$. The plot markers show experimental data, while the solid and dashed lines are the corresponding simulation results with and without additional fluctuations, respectively. Here, we compare the simulation and the experimental data at $T=2$ (left) and $T=3$ (right) Bloch periods. In the simulation we have assumed that about 9\% (left plot) and 12\% (right plot) of the fluorescence signal at each site contributes crosstalk to the fluorescence signal at each neighboring site.}
    \label{Fig:Supp_Bipartite}
\end{figure}

Finally, we use the TWA simulation to study the effect of experimental noise on the measurement of bipartite correlations shown in Fig~\ref{fig:trees}\,c. The results are summarized in Fig.~\ref{Fig:Supp_Bipartite}. We compare the experimental and simulation results at $T=2$ and $T=3$ Bloch periods of evolution time. If we consider no experimental imperfection in the simulation, the extracted bipartite correlations in the simulation show qualitative agreement with the experiment as a function of the parameter $s$. However, especially for low correlations in the simulation, the experimentally extracted correlations are significantly larger.

For a more accurate simulation of the experimental situation, we include two known experimental imperfections into the TWA simulation. Firstly, we include magnetic field fluctuations as explained before. In addition, we account for classical correlations between neighbouring ensembles due to the finite resolution of the fluorescence imaging. We expect that the magnetic field fluctuations lead to a reduction of correlations, while the imaging resolution increases the correlations especially in the Monna mapped case. Including these two effects in the simulation improves the agreement with the experimental data, especially for the Monna-mapped ordering.

In the experiment, there are still excess correlations in the physically ordered case for $s<0$. These may be due to decoherence effects, together with a slight difference in detection efficiency between the magnetic substates $\pm1$. Together these effects could explain the overall increase of correlation which we observe especially for the physical ordering of sites.

\normalem
\putbib[programmable]
\end{bibunit}